\documentstyle[12pt,equations]{article}
\begin{document}
\newcommand{\Ab}{\bar{A}}
\newcommand{\eb}{\bar{e}}
\newcommand{\et}{\tilde{e}}
\newcommand{\thone}{\theta_1}
\newcommand{\thtwo}{\theta_2}
\newcommand{\tcr}{T_{cr}}
\newcommand{\chit}{\tilde{\chi}}
\newcommand{\phit}{\tilde{\phi}}
\newcommand{\df}{\delta \phi}
\newcommand{\dr}{\delta \rho}
\newcommand{\dkl}{\delta \kappa_{\Lambda}}
\newcommand{\dkg}{\delta \kappa_{G}}
\newcommand{\dkcr}{\delta \kappa_{cr}}
\newcommand{\dxg}{\delta x_{G}}
\newcommand{\dx}{\delta x}
\newcommand{\lx}{\lambda}
\newcommand{\Lx}{\Lambda}
\newcommand{\ex}{\epsilon}
\newcommand{\ebs}{\bar{e}^2}
\newcommand{\ks}{k_s}
\newcommand{\gb}{\bar{g}}
\newcommand{\lb}{{\bar{\lambda}}}
\newcommand{\lbz}{\bar{\lambda}_0}
\newcommand{\lt}{\tilde{\lambda}}
\newcommand{\lr}{{\lambda}_R}
\newcommand{\lrt}{{\lambda}_R(T)}
\newcommand{\lbr}{{\bar{\lambda}}_R}
\newcommand{\lk}{{\lambda}(k)}
\newcommand{\lbk}{{\bar{\lambda}}(k)}
\newcommand{\lbkt}{{\bar{\lambda}}(k,T)}
\newcommand{\ltk}{\tilde{\lambda}(k)}
\newcommand{\mx}{{m}^2}
\newcommand{\mxk}{{m}^2(k)}
\newcommand{\mxkt}{{m}^2(k,T)}
\newcommand{\mb}{{\bar{m}}^2}
\newcommand{\mbb}{{\bar{m}}_2^2}
\newcommand{\mbu}{{\bar{\mu}}^2}
\newcommand{\mt}{\tilde{m}^2}
\newcommand{\mr}{{m}^2_R}
\newcommand{\mrt}{{m}^2_R(T)}
\newcommand{\mk}{{m}^2(k)}
\newcommand{\rht}{\tilde{\rho}}
\newcommand{\rhz}{\rho_{0R}}
\newcommand{\rhzt}{\rho_0(T)}
\newcommand{\rhztil}{\tilde{\rho}_0}
\newcommand{\rhzk}{\rho_0(k)}
\newcommand{\rhzkt}{\rho_0(k,T)}
\newcommand{\kx}{\kappa}
\newcommand{\kt}{\tilde{\kappa}}
\newcommand{\kk}{\kappa(k)}
\newcommand{\ktk}{\tilde{\kappa}(k)}
\newcommand{\Gammat}{\tilde{\Gamma}}
\newcommand{\Gammak}{\Gamma_k}
\newcommand{\wt}{\tilde{w}}
\newcommand{\be}{\begin{equation}}
\newcommand{\ee}{\end{equation}}
\newcommand{\een}{\end{subequations}}
\newcommand{\ben}{\begin{subequations}}
\newcommand{\beq}{\begin{eqalignno}}
\newcommand{\eeq}{\end{eqalignno}}
\pagestyle{empty}
\noindent
\begin{flushright}
CERN-TH/96-190
\end{flushright} 
\vspace{3cm}
\begin{center}
{{ \Large  \bf
The Electroweak Phase Transition \\
through the Renormalization Group
}}\\
\vspace{10mm}
N. Tetradis 
\vspace {0.5cm}
 \\
{\em 
CERN, Theory Division, \\
CH-1211, Geneva 23, Switzerland
\footnote{E-mail: tetradis@mail.cern.ch.} 
} 

\end{center}

\setlength{\baselineskip}{20pt}
\setlength{\textwidth}{13cm}
 
\vspace{3.cm}
\begin{abstract}
{We study the high-temperature phase transitions for the 
Abelian and $SU(2)$ Higgs models, using 
the exact renormalization group. The evolution
equation for a properly-defined coarse-grained 
free energy is solved. 
The phase diagram of the Abelian Higgs model has a region of
second-order phase transitions governed by fixed points, for 
which we calculate critical exponents and crossover curves.
It also has a region of first-order phase transitions, which we
discuss in terms of the coarse-grained free energy. The $SU(2)$ Higgs model
for small Higgs field masses has only first-order phase transitions. 
We derive their properties and compare with lattice and perturbative results.
For larger Higgs field masses the strength of the first-order phase transitions
diminishes. For masses larger than 80-100 GeV we find evidence that
they are replaced by analytical crossovers.}
\end{abstract}
\vspace{1cm}
\noindent
CERN-TH/96-190 \\
August 1996
\clearpage
\setlength{\baselineskip}{15pt}
\setlength{\textwidth}{16cm}
\pagestyle{plain}
\setcounter{page}{1}

\newpage

\setcounter{equation}{0}
\renewcommand{\theequation}{{\bf 1.}\arabic{equation}}

\section*{1. Introduction}

The restoration of the electroweak symmetry 
at temperatures of the order of 100 GeV 
is an important element of the Big Bang cosmology
\cite{original}. Its most prominent observable 
consequence is related to the possibility of 
generating the baryon asymmetry of the Universe 
during the electroweak phase transition \cite{baryon}. 
The precise determination of the created baryon number,
which could be compared with the observed matter density of the 
Universe, depends very sensitively on the details of the
phase transition. This fact has instigated numerous 
studies of its characteristics during the last few years. 
(For recent reviews 
with an extensive list of references, see refs. \cite{review,jansen}.)
Most of the older calculations employed 
the perturbative approach for the determination of 
the temperature-dependent effective potential \cite{pert},
from which the properties of the phase transition 
can be inferred.  
This approach has been pursued up 
to two loops \cite{twoloop}.
However, it has been known for a long time \cite{gaugediv} 
that the perturbative expansion breaks 
down near and in the symmetric phase of 
gauge theories. The reason is that 
at non-zero temperature the relevant expansion parameter is
$\bar{e}^2 T / m_A(T)$. Here $\bar{e}$ is the bare gauge coupling,
and $m_A(T)$ the total thermal mass of the gauge field, accounting
for possible thermal screening effects. 
The problem is due to the fact that no thermal 
corrections are expected for 
the mass of the transverse components of the gauge field within
perturbation theory.  
This causes the perturbative expansion for the electroweak theory 
to diverge for small Higgs field expectation
values, for which the zero-temperature mass of the gauge field is small.
In order to overcome this difficulty, alternative approaches have
been followed. Gap equations (truncated versions of Schwinger-Dyson
equations) \cite{buchgap,owe}
have been employed in order to 
obtain systematic resummations of infinite subclasses
of perturbative contributions.
The $\ex$-expansion \cite{arnold} has also been used in order to
obtain insight into the non-perturbative character of the phase transition. 
The most reliable quantitative results have been obtained through the
lattice approach \cite{ilg,kaj,fodor}. However, the underlying 
dynamics which results in a certain physical behaviour is often
obscured by the Monte-Carlo simulations. The analytical approaches
offer a more intuitive understanding. Also, the requirement of 
long computer time for the simulations means that 
the exploration of the full phase diagram for a particular
model is often a formidable task. 

The purpose of this paper is to present a different approach based on the
exact renormalization group \cite{wilson}. We employ the method of the 
effective average action $\Gamma_k$ \cite{averact}, which is
a generalization in the continuum   
of the blockspin action \cite{kadanoff}.
It results from the integration of fluctuations with characteristic
momenta larger than a given scale $k$ (corresponding to the inverse of the
blockspin size). The scale $k$ acts as an effective
infrared regulator and gives us control over the regions
in momentum space from which divergences are expected to arise in 
perturbation theory. The dependence of $\Gamma_k$ on $k$ is described
by an exact renormalization-group equation \cite{exact,gauge}, typical
of the Wilson approach to the renormalization group \cite{wilson}
\footnote{
For other versions of exact renormalization-group equations 
see ref. \cite{reneq}, and for related analyses of such equations
see ref. \cite{related}.
}.
For large values of $k$ (of the order of the ultraviolet cutoff 
$\Lx$ of the theory) 
the effective average action is equal to the classical action 
(no fluctuations are integrated out), while for $k \rightarrow 0$
it becomes the standard effective action (all fluctuations are 
integrated out). 
As a result, the solution
of the exact renormalization-group equation, with the classical
action as initial condition, gives all the physically relevant information
for the renormalized theory at low scales. The formalism is 
constructed in Euclidean space, which makes the consideration of
temperature effects straightforward, through the imposition of periodic
boundary conditions in the time direction.  
This approach has been used in the past for the discussion of the
phase transition for the $O(N)$-symmetric pure scalar theory 
\cite{trans,largen,indices}. 
The perturbative study of the
phase transition for this theory runs into infrared 
problems very similar to those for the gauge theory.
Higher orders in the perturbative expansion involve powers of the
quantity 
$\bar{\lx} T / m_s(T)$ (with $\bar{\lx}$ the bare quartic coupling and 
$m_s(T)$ the temperature-dependent mass), 
which diverges near the critical temperature. 
This problem was resolved in refs. 
\cite{trans,largen,indices}.
The evolution of the 
potential as a function of $k$ was studied, using
a polynomial approximation for its dependence on the field 
expectation value. It was shown that 
there are no divergent quantities
in the true physical picture. 
The temperature-dependent quartic coupling $\lx(T)$ is strongly
renormalized near the critical temperature, so that quantities such as
${\lx}(T) T / m_s(T)$ always stay finite. 
A detailed picture of the second-order phase transition 
emerged, with a complete mapping of the phase diagram, calculation of 
critical temperature and exponents, etc. 
The same method was subsequently applied to two-scalar theories
\cite{twoscalar,crossover,twoscalar4d}, where again it gave
a reliable, detailed picture of the more complicated phase
diagram, with first-order and second-order phase transitions, crossover
phenomena, tricritical points, critical exponents, etc. 
The assumption for a polynomial dependence of the potential 
on the field expectation value was relaxed in ref. \cite{num}.
The numerical integration of the full partial 
differential equation, for the dependence on both the field
and the scale $k$, was achieved through appropriate algorithms. 
An immediate consequence of this development was the 
calculation of the equation of state for the pure scalar theory
\cite{eos}.
The results were confirmed by fully analytical solutions
in the large $N$ limit \cite{analy}. 
The framework for the study of gauge theories has been provided by
the generalization of the method of the effective average action
and the derivation of appropriate 
exact renormalization-group equations \cite{gauge,abelian,runngauge}.
The study of the four-dimensional Abelian Higgs model 
\cite{abelian4d}
has
reproduced the Coleman-Weinberg mechanism for radiative
symmetry breaking \cite{colwein}. The same model in three-dimensions
was studied in ref. \cite{supercond}, where a polynomial
form of the potential was assumed.
The phase diagram and the phase transitions, which are  
relevant for the 
superconductor phase transition, were discussed in detail.
Also the high-temperature phase transition for the
$SU(2)$ Higgs model was discussed in ref. \cite{bastian}. 
The potential was calculated through an evolution
equation, in which the effective infrared cutoff $k$ was
replaced by the Higgs field expectation value. 

In this paper we discuss the Abelian and $SU(2)$ Higgs 
models at non-zero temperature, by starting from the 
exact renormalization-group equation for the effective average action.
We relax the polynomial approximation of ref. \cite{supercond}
and keep both the scale 
$k$ and the Higgs field expectation value as independent variables.
Bringing the renormalization-group equation into a manageable form
that permits its solution
requires various approximations, which 
we spell out in detail.  
We list the resulting sources of errors and
estimate their magnitude, whenever possible. 
However, some of the errors are difficult to estimate. 
The largest source of uncertainty is introduced 
by our assumption for the invariants appearing in  
the effective average action. 
Our ansatz is a truncated expression
that neglects higher derivative terms.
A quantitative estimate of the induced error 
requires the comparison with more extended truncations.
As a result, an intrinsic check of the 
accuracy of our conclusions can be obtained through extensions
of this work, which incorporate more invariants in a systematic way.
First attempts in this direction were made in refs. \cite{indices,system}
for scalar theories. 
An alternative way of checking our results is to compare them with 
the results of other approaches, such as perturbation theory or
lattice calculations.
We make these comparisons in the following sections. 
Despite the limitations in accuracy, our approach 
provides a complete and detailed picture of the phase 
structure of the Abelian and $SU(2)$ Higgs models. 
We discuss the phase diagram in terms of the 
effective average potential, as a function of the Higgs field expectation
value and the running scale $k$. As the potential is the
generating functional of all 1PI Green's functions at zero external momenta, 
our solutions provide information for the running of all the generalized
couplings of the theory. The discussion of 
fixed points, which govern the second-order phase transitions, 
is carried out with the inclusion of both relevant and 
irrelevant parameters (apart from the higher derivative terms in the action). 
Moreover, the consideration of the full potential 
permits a reliable study of first-order phase transitions. The 
appearance of new minima and their relative stability
can be investigated in detail, without 
extrapolations from the form of the potential at the original
minimum \cite{supercond}, or reliance on perturbation theory. 
As a result, we are able to build the complete 
phase diagram of the Abelian and $SU(2)$ Higgs models. 

Before presenting our analysis, we mention two important points, which
will be encountered in the following: \\
a) The discussion 
of a first-order phase transition usually relies on the study
of a non-convex potential. The decay of unstable minima 
is associated either with tunnelling fluctuations through 
barriers or thermal fluctuations above them. However, the 
effective potential is expected to be a convex quantity, with no
such barriers. The resolution of this paradox lies in the 
realization that the effective potential is convex because
the tunnelling or thermal fluctuations are incorporated in it
\cite{convex}. They are associated with the low-frequency modes
which are integrated out at the later stages of the 
evolution of a quantity such as the effective average potential. 
A natural approach for the study of first-order phase transitions
would be to separate the problem in two parts. First, the
high-frequency modes are integrated out, with the possible
generation of new mimima through radiative symmetry breaking. 
Subsequently, the decay of unstable minima is discussed with
semiclassical techniques, in the non-convex potential that has 
resulted from the first step. 
This leads us to the notion of the coarse-grained free energy, which
is fundamental in statistical physics. Every physical system
has a characteristic length scale associated with it. The dynamics
of smaller length scales is integrated out, and
is incorporated in the parameters of the 
free energy one uses for the study of the behaviour at larger scales.
The Wilson approach to the renormalization group 
provides the realization of this intuitive idea. 
The coarse-graining scale 
(corresponding to the inverse of the characteristic length scale of the 
system) can be identified with $k$. The potential at this non-zero scale is 
not necessarily convex. In this paper we provide an explicit 
demonstration of how such a potential can 
be obtained starting from the classical action of a field theory.  
We also find that the intuitive separation of the problem in two parts
is natural for strongly first-order phase transitions, but not so
for weakly first-order ones. All this is discussed in detail
in section 8. The notion of coarse graining is 
absent in perturbation theory. This is the main reason 
for the non-convergence of the perturbative series near the 
maxima of the classical potential, and the appearance of 
imaginary parts in the perturbative effective potential. \\
b) The infrared problem near and in the symmetric phase of gauge theories
is more complicated than for scalar theories.
The divergence of the quantity $\bar{e}^2 T / m_A(T)$ in the perturbative
expansion for gauge theories is reminiscent of the divergence of  
$\bar{\lx} T / m_s(T)$ for scalar theories. As we have mentioned, 
in the latter case the renormalization group approach resolves the
problem, through the identification of the relevant dynamics which is
associated with infrared attractive fixed points. 
A similar behaviour will be observed for the Abelian Higgs model in
section 7. 
However, an additional complication is expected for 
asymptotically free theories, such as the $SU(2)$ Higgs model.
The absence of an effective infrared cutoff for the gauge field fluctuations
near and in the symmetric phase leads to the increase of the renormalized
gauge coupling. When this exceeds a critical value, the emergence of a
confining regime is observed \cite{symm}. This behaviour 
is expected near the origin of the potential, where the zero-temperature
mass of the gauge field is small, and for small values of the
coarse-graining scale $k$. In this region the relevant observables 
are associated with bound states. An appropriate parametrization of 
the effective average action in terms of composite operators 
must be introduced, which is more 
efficient in capturing the relevant dynamics \cite{bound}. 
In this work we do not consider in detail this last part of the evolution,
which requires an extensive analysis of its own. 
Instead we follow the evolution
within the parametrization of the effective average action in terms of 
fundamental fields, up to the point of emergence of the confining regime. 
The most important element of the last stage of the evolution, for our 
discussion,
is the appearance of an expectation value for the operator $F^2$, with 
a negative contribution to the free energy density \cite{condensate}. 
This contribution modifies the
effective average potential, as it appears only near its origin. 
We approximate it by $-(c k_{conf})^3 T$, where
$k_{conf}$ is the point in the evolution where confinement sets in, and
$c$ is of order one. The detailed analysis is given in section 10.

The paper is organized as follows:
In section 2, we derive the evolution equation for the potential in the
Abelian Higgs model. Two derivations are presented: a rather crude but 
intuitive one, and a more rigorous one, with explicit discussion of the 
approximations introduced. In section 3, the evolution equations for the
potential and the gauge coupling are cast in a form that does not depend on
the running scale $k$. This makes the identification of fixed points
straightforward.
The non-zero-temperature formalism is introduced in section 4, 
and the dimensional reduction of the four-dimensional, 
non-zero-temperature theory
to an effective 
three-dimensional, zero-temperature one is derived in section 5.
In section 6, the problem is reformulated as a three-dimensional one, with
an appropriate initial condition. In section 7, the phase diagram for the
Abelian Higgs model is derived in terms of the effective 
average potential, which depends on the Higgs field expectation value
and the running scale $k$. Fixed points and tricritical points are identified,
which govern second-order phase transitions. 
The critical behaviour is discussed in terms of critical  exponents and
crossover curves. In section 8, the first-order phase transitions are studied.
The notion of coarse graining is introduced, and a non-convex potential 
is derived, which can be employed for the study of tunnelling.
Weakly first-order phase transitions are shown to be difficult to describe
in terms of semiclassical estimates for the nucleation probability.
In section 9, the formalism is developed for the study of the 
$SU(2)$ Higgs model. The presentation is brief, because the 
derivations are straightforward generalizations 
of those in sections 2--6.
The phase transitions for the $SU(2)$ Higgs model are discussed in
section 10. Arguments are presented on how the first-order phase
transitions turn to continuous crossovers for large masses
of the Higgs field. Our conclusions are given in section 11.

\setcounter{equation}{0}
\renewcommand{\theequation}{{\bf 2.}\arabic{equation}}

\section*{2. Evolution equation for the potential in the Abelian Higgs
model}

Before presenting the rigorous derivation, 
it is instructive to derive the evolution equation for the 
potential based on an intuitive argument, along the lines of
refs. \cite{averact,trans}. 
The one-loop effective potential
for the Abelian Higgs model with
$N_c$ complex scalars, in $d$ dimensions, in the Landau gauge, is
given by the expression
\beq
U^{(1)}_k(\rho) = V(\rho)
+ \frac{1}{2} (2 \pi)^{-d}
\int d^dq~
&\left\{ 
\ln \left[ P(q) + V'(\rho) + 2 V''(\rho) \rho \right] 
\rule{0mm}{5mm} \right.
\nonumber \\
&~+ (2N_c-1) \ln \left[ P(q) + V'(\rho) \right]
\nonumber \\
&\left. \rule{0mm}{5mm} 
+ (d-1) \ln \left[ P(q) + 2 \bar{e}^2 \rho) \right]
\right\},
\label{twoone} \eeq
where 
$V(\rho)$ is the classical potential and 
$\bar{e}^2$ the bare gauge coupling. 
The three terms in the r.h.s. are the contributions of the 
radial scalar mode, the Goldstone modes and the gauge field.
We have defined the variable 
\be
\rho = \frac{1}{2} \phi^a \phi_a,
\label{twotwo} \ee
which we 
frequently use in the following. 
Primes denote derivatives with respect to $\rho$: 
$V'(\rho) =  {d V}/{d \rho}$.
We assume that the above integral is regulated by an ultraviolet
cutoff $\Lx$.
The inverse propagator $P(q)$ is given by 
$P(q)=q^2$ in perturbation theory. 
We would like to introduce an effective cutoff $k$ for the 
low frequency modes, so that the momentum integration in
eq. (\ref{twoone}) does not receive contributions from modes with 
characteristic momenta $q^2 \ll k^2$.
This can be achieved through the following modification of the 
inverse propagator:
\beq
P(q) = &\frac{q^2}{1-f^2_k(q)}
\nonumber \\
f^2_k(q)= &\exp \left( -\frac{q^2}{k^2} \right). 
\label{twothree} \eeq
This form of $P$ provides for an infrared cutoff, which acts like 
a mass term $\sim k^2$ for the modes with $q^2 \ll k^2$, while
it leaves unaffected the modes with $q^2 \gg k^2$. The potential
now depends on $k$, as indicated by the subscript in eq. (\ref{twoone}). 
We would like to derive an evolution equation for the change of 
$U_k$ with the scale $k$ and follow the evolution for 
$k \rightarrow 0$. This will give us control over the 
region of the momentum integration from which infrared divergences 
are expected to arise in perturbation theory. 
For this purpose we take the logarithmic derivative
with respect to $k$ 
and substitute $U_k$ for $V$ and the running gauge coupling 
$e^2(k)$ for the bare one $\bar{e}^2$
in the r.h.s. of eq. (\ref{twoone}). The
intuitive justification for this replacement is based on the fact 
that the new contributions to the momentum integration, when $k$ is
lowered by a small amount $\Delta k$, come from the region 
$k- \Delta k < q < k$. The relevant mass terms and couplings which should
appear in the evolution equation are the running ones at the 
scale $k$ (which for the scalar modes are related to derivatives of $U_k$)
and not the bare ones. This ``renormalization-group improvement''
results in the evolution equation 
\beq
\frac{\partial U_k(\rho)}{\partial t} = \frac{1}{2} (2 \pi)^{-d}
\int d^dq~ \frac{\partial P}{\partial t}
&
\left( 
\frac{1}{P(q) + U_k'(\rho) + 2 U_k''(\rho) \rho} 
+ \frac{2N_c-1}{P(q) + U_k'(\rho)} 
\rule{0mm}{5mm} \right.
\nonumber \\
& \left.
+ \frac{d-1}{P(q) + 2 e^2(k) \rho}
\rule{0mm}{5mm} \right),
\label{twofour} \eeq
where
$t = \ln (k/\Lx)$, with $\Lx$ identified with
the ultraviolet cutoff of the theory.
The above equation must be supplemented with an equation for the
running of $e^2$
\be
\frac{d e^2(k)}{dt} = ~\beta_{e^2}.
\label{twosix} \ee
For $k=\Lx$ the infrared and ultraviolet cutoffs coincide, and no
integration of fluctuations has taken place. This determines the 
initial conditions for the solution of eqs. (\ref{twofour}), (\ref{twosix})
as $U_{\Lx}(\rho) = V(\rho)$, $e^2(\Lx)=\bar{e}^2$.
In the opposite limit $k \rightarrow 0$ one 
recovers the effective potential $U(\rho) \equiv U_0(\rho)$
and the renormalized gauge coupling 
$e^2_R \equiv e^2(0)$.
In this heuristic derivation we have not addressed
the problem of reconciling gauge invariance and the presence of
a cutoff. This question can be resolved in the context of a 
rigorous approach, which we describe in the next paragraph. 

A rigorous derivation of eq. (\ref{twofour}), with an explicit 
determination of the approximations involved, is given in 
refs. \cite{gauge,abelian}. We give here only a brief sketch of the
derivation, which employs the background-field formalism. 
One starts with a gauge-invariant classical action 
$S(\chi,\bar{A}+a)$ for a complex,
$N_c$-component, scalar field
$\chi$ and fluctuations $a_{\mu}$ of the gauge field 
around some arbitrarily chosen background field $\bar{A}_{\mu}$.
To this a gauge-fixing term 
$S_{gf}={1}/{(2 \alpha)} \int d^d x 
\left( \partial^{\mu} a_{\mu} \right)^2$
is added. Additional gauge-invariant 
terms $\Delta_k S_S$, $\Delta_k S_G$ 
are constructed, which act as effective infrared cutoffs for the
scalar and gauge fields respectively. Their detailed form 
is given in refs. \cite{gauge,abelian}
and involves a cutoff function $R_k(q)$
closely related to the form of the 
effective inverse propagator of eqs. (\ref{twothree}). 
Through the standard Legendre transformation and 
after the removal of the infrared cutoff terms, one obtains 
the effective average action $\Gamma_k(\phi,A,\bar{A})$. This action is
invariant under simultaneous gauge transformations of $\phi$ and 
both gauge fields $A$ and $\bar{A}$. 
Physical observables computed from it 
should be independent of $\bar{A}$ and the gauge parameter $\alpha$. 
$\Gamma_k(\phi,A,\bar{A})$ includes the 
effects of the integration of modes with 
$q^2 > k^2$. For $k=\Lx$ it coincides with the 
classical action plus the gauge-fixing term, while for 
$k \rightarrow 0$ it reproduces the effective action 
in the background-field formalism.
It obeys the exact renormalization-group equation
\be
\frac{\partial \Gamma_k(\phi,A,\bar{A})}{\partial t} = ~\frac{1}{2}
{\rm Tr} \left\lbrace
\left( \Gamma^{(2)}_k(\phi,A,\bar{A}) + R_k \right)^{-1} 
\frac{\partial R_k}{\partial t}
\right\rbrace,
\label{twoseven} \ee
with $\Gamma^{(2)}_k$ the matrix of second functional derivatives
with respect to $\phi$ and $A$ at fixed $\bar{A}$.
The solution of the above equation is 
a difficult task, because the effective average action
involves an infinite number of terms, even if the classical action
has only a few. 
The manifest preservation of the gauge symmetry in the formulation
constrains the form of $\Gamma_k(\phi,A,\bar{A})$. However, 
a truncation is unavoidable in practical applications. We approximate the
effective average action by 
\beq
\Gamma_k(\phi,A,\Ab) = 
\int d^dx~
&\left\{ 
\left( D^{\mu}\phi_a \right)^*
\left( D_{\mu}\phi^a \right)
+ U_k(\rho) + \frac{1}{4} Z_{F,k} F_{\mu \nu} F^{\mu \nu}
\rule{0mm}{5mm} \right.
\nonumber \\
&~\left. \rule{0mm}{5mm} 
+ \frac{1}{2 \alpha} 
\left[ \partial^{\mu} \left( A_{\mu}-\Ab_{\mu} \right) \right]^2
\right\},
\label{twoeight} \eeq
with $D_{\mu}=\partial_{\mu}+i \eb A_{\mu}$, $\eb$ the bare 
gauge coupling and $\rho$ given by eq. (\ref{twotwo}).
The resulting evolution equation for the effective average
potential, for the gauge
parameter $\alpha \rightarrow 0$, is given by
eq. (\ref{twofour}),
with the running gauge coupling defined according to
\be
e^2(k)= Z^{-1}_{F,k} \eb^2.
\label{twonine} \ee

Some remarks are due at this point, concerning the nature of 
our approximations. 
In eq. (\ref{twoeight}) we have neglected 
the wave-function renormalization of the 
scalar field. This is expected to be a good approximation, 
because the anomalous dimension of the three-dimensional theory, 
which is relevant near the phase transition, is small 
($\sim$ 10\% or less) and 
gives only small quantitative corrections \cite{supercond}.
The higher derivative terms involving the scalar field $\phi$
are not expected to introduce significant 
corrections either. A solution of the evolution equation
for the three-dimensional pure scalar theory at the same 
truncation level leads to
an accurate determination of the equation of state
\cite{eos}. The largest uncertainties in our approach stem
from the neglected terms involving the gauge fields 
$A$ and $\Ab$. 
In particular, the assumption that the only
$\Ab$ dependence comes through a standard gauge-fixing term 
is rather crude. 
The $\bar{A}$ dependence of $\Gamma_k(\phi,A,\bar{A})$ 
is constrained by exact identities \cite{gauge,abelian}
(which include the generalized Ward identities \cite{filip}).
In general one can write
\be
\Gamma_{k}(\phi,A,\Ab) = \bar{\Gamma}_k(\phi,A)
+\frac{1}{2 \alpha} \int d^d x 
\left[ \partial^{\mu} \left( A_{\mu}-\bar{A}_{\mu} \right) \right]^2
+\hat{\Gamma}_k(\phi,A,\bar{A}).
\label{twoten} \ee
In our truncation we have approximated the 
gauge-invariant part $\bar{\Gamma}_k(\phi,A)$
through the first three terms in the r.h.s. of 
eq. (\ref{twoeight}) and completely neglected 
the piece
$\hat{\Gamma}_k(\phi,A,\bar{A})$.
This seems reasonable as a first step, as 
we are mainly interested
in the evolution of $U_k(\rho)$ and $e^2(k)$.
However, for general $k$ (different from $\Lx$ or 0),
our assumption violates the exact identities
governing the dependence on the background field. 
A partial compensation for the neglected terms 
can be obtained through the introduction of 
appropriate correction factors in the evolution of 
the invariants in $\bar{\Gamma}_k(\phi,A)$ 
\cite{abelian}. These correction factors mainly influence the 
evolution of the gauge coupling and are given in the following 
section 
\footnote{In an alternative approach, explicit gauge invariance
is not imposed for the action with a cutoff \cite{ellwanger}. 
The consistency of the action when the cutoff is removed is guaranteed
by making sure that generalized Ward identities 
are satisfied by the ansatz during the whole evolution.}. 
A quantitative estimate of the error 
induced by the truncated form of the effective average action
requires comparison with more extended truncations, and is 
beyond the scope of this work. However, we shall be able to
draw conclusions about the accuracy of our results through a 
comparison with the results of other approaches.

\setcounter{equation}{0}
\renewcommand{\theequation}{{\bf 3.}\arabic{equation}}

\section*{3. Scale-invariant form of the evolution equations}

It is convenient to cast the evolution equation (\ref{twofour}) in 
a form that does not depend explicitly on the scale $k$.
This makes the identification of possible fixed points easier.
For this reason we define the dimensionless quantities
\beq
u_k(\rht) = &~k^{-d} U_k(\rho)
\nonumber \\
\rht = &~k^{2-d} \rho
\nonumber \\
{\et}^2(k) = &~k^{d-4} e^2  (k)
= k^{d-4} Z_{F,k}^{-1} \bar{e}^2.
\label{threeone} \eeq
Primes on $u_k$ denote derivatives with respect to $\rht$.
We can now rewrite eq. (\ref{twofour}) as 
\beq
\frac{\partial u_k}{\partial t} = & 
~-d u_k +(d-2) \rht u'_k
\nonumber \\
&~-v_d L^d_0 \left( u'_k +2 u''_k \rht \right)
-(2N_c-1) v_d L^d_0 \left( u'_k \right)
-(d-1) v_d L^d_0 \left( 2 \et^2 \rht \right),
\label{threetwo} \eeq
where
\be
v_d^{-1}=2^{d+1} \pi^{d/2} \Gamma \left( \frac{d}{2} \right).
\label{twofive} \ee 
In the second line we recognize the contributions of the
radial  scalar mode, the Goldstone modes and the gauge field.
In the following we shall find it more convenient to integrate 
numerically the evolution equation for $u'_k$. This reads
\beq
\frac{\partial u'_k}{\partial t} = & 
~-2 u'_k +(d-2) \rht u''_k
\nonumber \\
&~+v_d (3 u''_k +2 \rht u'''_k) L^d_1 \left( u'_k +2 u''_k \rht \right)
+(2N_c-1) v_d u''_k L^d_1 \left( u'_k \right)
\nonumber \\
&~+ 2 (d-1) v_d \et^2 L^d_1 \left( 2 \et^2 \rht \right).
\label{threethree} \eeq

The dimensionless functions $L^d_n(w)$ 
in the above equations
are given by 
\beq
L^d_n(w) = &
- n k^{2n-d} \pi^{-\frac{d}{2}} \Gamma \left( \frac{d}{2} \right) 
\int d^d q \frac{\partial P}{ \partial t} (P + w)^{-(n+1)} \nonumber \\
= &
- n k^{2n-d} 
\int_0^{\infty} dx x^{\frac{d}{2}-1} 
\frac{\partial P}{ \partial t} (P + w)^{-(n+1)},
\label{threefour} \eeq
with $x=q^2$ and
$P$ defined in eqs. (\ref{twothree}).
They have been discussed extensively in refs. 
\cite{averact,indices,convex} 
(for various forms of the infrared-regulating function).
Their most interesting property, for our discussion, is that  
they fall off for large values of $w$, following a power law. As 
a result they introduce a threshold behaviour for the 
contributions of massive modes to the evolution equations. 
The various contributions to the evolution equations 
involve $L^d_n$ integrals with the mass eigenvalues divided by $k^2$
as their arguments.
When the running squared 
mass of a massive mode 
becomes much larger than the scale $k^2$ 
(at which the system is probed), these
contributions vanish and the massive modes decouple. 
We evaluate the integrals $L^d_n(w)$ 
numerically and use numerical fits for the solution of 
the evolution equations.

The last remaining element that we need is the evolution equation 
for the gauge coupling. For the truncation that we are using,
this equation 
can be inferred from the results of ref. \cite{abelian} and reads 
\be
\frac{d \et^2}{dt} = 
(d-4)\et^2 + \frac{4}{3} v_d N_c \et^4 l^d_{gc}.
\label{threefive} \ee
The constant $l^d_{gc}$ incorporates a correction factor
which partially compensates for the crude treatment of the 
background-field dependence in our truncation
(see discussion at the end of the last section).
Its values in four and three dimensions are
\be
l^4_{gc}=1,~~~~~~~~~~~~~~~~~~~~~~~~~~l^3_{gc}=0.844.
\label{threesix} \ee
Notice that our truncation neglects the $\rho$ dependence of the
gauge coupling. This is the reason why the threshold functions,
which normally would appear in the r.h.s. of eq. (\ref{threefive})
\cite{abelian},
have been set equal to their value for zero argument. This means 
that the running masses of the various modes have been
approximated by zero in eq. (\ref{threefive}),
and no threshold effects are expected 
in the evolution of the gauge coupling. However, the 
threshold effects in the evolution of the potential are 
fully accounted for. The threshold behaviour in the evolution of 
the gauge coupling will
be important for our discussion of the phase transition for the 
$SU(2)$ Higgs model. An appropriate modification to the evolution
equation will be introduced at that stage.

In the following sections we describe numerical 
solutions of the coupled system of equations 
(\ref{threethree}), (\ref{threefive}). We explicitly
solve eq. (\ref{threethree}) as a
partial differential equation, making no additional approximations.
An alternative approach has been followed in ref. \cite{supercond}.
There, eq. (\ref{threethree}) is transformed into an infinite system
of ordinary differential equations for the minimum of the potential 
and the coefficients of the Taylor expansion of the potential around it.
This infinite system is truncated at a finite number of equations, which 
amounts to a polynomial approximation for the potential. 
For example, if the potential is approximated by a quartic polynomial and 
threshold effects are neglected, eq. (\ref{threethree}) gives
\beq
\frac{d \kx}{dt} = &(2-d) \kx + 4 (N_c+1) v_d l^d_1 
+ 4 (d-1) v_d l^d_1 \frac{\et^2}{\lt} 
\label{threeseven} \\
\frac{d \lt}{dt} = &(d-4) \lt + 4 (N_c+4) v_d l^d_2 \lt^2
+ 8 (d-1) v_d l^d_2 \et^4,
\label{threeeight} \eeq
with $l^d_n$ constants of order one,
for the minimum $\kx(k)$ of the rescaled potential 
and quartic coupling $\lx(k)=u''_k(\kx)$.
The disadvantage of
this approach lies in the imprecise treatment of logarithmic contributions
to the potential, and the coarse description of the cases when a 
second minimum appears in the potential, 
leading to a first-order phase transition. Notice, however, that
the effect of the wave-function renormalization of the 
scalar field 
and threshold effects in the running of the 
gauge coupling have been taken into account in \cite{supercond}, while 
we have neglected them here.  
It is interesting also to compare with 
the leading result from the $\ex$-expansion
for the evolution of $\lx$ and $\et^2$ \cite{ginsparg,russell,arnold} 
\beq
\frac{d \lt}{d t} = &-\ex \lt
+ 4 (N_c+4) v_4  \lt^2
+ 24 v_4  \et^4
- 24 v_4  \et^2 \lt
\label{threenine} \\
\frac{d \et^2}{d t} = &-\ex \et^2 +\frac{4}{3} v_4 N_c \et^4.
\label{threeten} \eeq
It is clear that, for small $\ex=4-d$, 
eqs. (\ref{threenine}), (\ref{threeten})
are in agreement with eqs. (\ref{threeeight}), (\ref{threefive}) respectively
($l^4_1=l^4_2=1$),
apart from the last term in the r.h.s. of eqs. (\ref{threenine}), which
we could have also reproduced by taking into account the
wave-function renormalization of the scalar field \cite{supercond}.
However, significant deviations in the values of the numerical coefficients 
are observed for $\ex=1$.

\setcounter{equation}{0}
\renewcommand{\theequation}{{\bf 4.}\arabic{equation}}

\section*{4. The non-zero-temperature formalism}

In order to extend the formalism of the 
previous sections to non-zero temperature 
we need to recall that, 
in Euclidean formalism, non-zero-temperature $T$ results 
in periodic boundary conditions in the time direction 
(for bosonic fields),
with periodicity $1/T$ \cite{kapusta}.
This leads to a discrete spectrum for the zero component of the momentum
$q_0$ 
\be
q_0 \rightarrow 2 \pi m T,~~~~~~~~~m=0,\pm1,\pm2,...
\label{fourone} \ee
As a consequence the integration over $q_0$ is replaced by 
a summation over the
discrete spectrum
\be
\int \frac{d^d q}{(2 \pi)^d} \rightarrow 
T \sum_m \int \frac{d^{d-1}\vec{q}}{(2 \pi)^{d-1}}. 
\label{fourtwo} \ee
Another important point is that, due to the explicit breaking of
Lorentz invariance by the temperature, the $A^0$ component of the 
gauge field evolves differently from the other three components. 
Therefore, it must be treated as a separate scalar 
in the truncated action of eq. (\ref{twoeight}), which develops its own 
mass term $m^2_{A^0}(k,\rho,T)$. 
With the above remarks in mind we can generalize our master equation 
(\ref{twofour}) in order to take into account the temperature effects. 
For the temperature-dependent effective 
average potential $U_k(\rho,T)$ 
we obtain
\beq
\frac{\partial U_k(\rho,T)}{\partial t} &= 
\frac{1}{2} (2 \pi)^{-(d-1)} T \sum_m \int d^{d-1} \vec{q}~~ 
 \frac{\partial P}{\partial t} \times
\left( 
\frac{1}{P(q) + U_k'(\rho,T) + 2 U_k''(\rho,T) \rho} 
~\right. \rule{0mm}{5mm} 
\nonumber \\
&
\rule{0mm}{5mm} \left. 
+ \frac{2N_c-1}{P(q) + U_k'(\rho,T)} 
+ \frac{d-2}{P(q) + 2 e^2(k,T) \rho}
+ \frac{1}{P(q) + m^2_{A^0}(k,\rho,T)}
\right),
\label{fourthree} \eeq
with the implicit replacement 
\be
q^2 \rightarrow \vec{q}^2 + 4 \pi^2 m^2 T^2
\label{fourfour} \ee
in eqs. (\ref{twothree})
for $P$. 
For $k \sim \Lx \gg T$,
with $\Lx$ the ultraviolet cutoff, 
the thermal corrections are negligible (see next paragraph).
As a result the initial condition for the solution of eq. (\ref{fourthree})
is the same as for the zero-temperature case:
$U_\Lx(\rho,T)= V(\rho)$, with $V(\rho)$ the classical potential. 
The temperature-dependent effective potential
\cite{doljac}--\cite{linde}
is obtained from $U_k(\rho,T)$ in the limit $k \rightarrow 0$. 

We can put 
eq. (\ref{fourthree}) in a scale-invariant form analogous to 
eq. (\ref{threetwo}), by absorbing all the temperature 
dependence in generalized 
$L^d_n$ functions. They read
\be
L^d_n(w,T) = 
- 2 n k^{2n-d} \pi^{-\frac{d}{2}+1} \Gamma \left( \frac{d}{2} \right) 
 T \sum_m
\int d^{d-1} \vec{q}~~ \frac{\partial P}{ \partial t} (P + w)^{-(n+1)}, 
\label{fourfive} \ee
where again the replacement 
(\ref{fourfour}) is assumed in $P$.
Their basic properties can be established
analytically. 
For $T \ll k$ the summation over discrete 
values of $m$ in the expression (\ref{fourfive}) is equal to the 
integration over a continuous range of $q_0$ up to exponentially small 
corrections.
Therefore 
\be
L^d_n(w,T) = L^d_n(w)~~~~~~~~~{\rm for}~~T \ll k.
\label{foursix} \ee
In the opposite limit $T \gg k$ the summation over $m$ is
dominated by the $m=0$ contribution. 
Terms with non-zero values of $m$ are suppressed    
by $  \sim \exp \left( -( mT/k)^{2 } \right).$
The leading contribution results in 
the simple expression
\be 
L^d_n(w,T) = \frac{v_{d-1}}{v_d} \frac{T}{k} L^{d-1}_n(w)~~~~~~{\rm 
for}~~T \gg k,
\label{fourseven} \ee
with $v_d$ defined in eq. (\ref{twofive}).
The two regions of $T/k$ in which $L^d_n(w,T)$ is given by the 
equations (\ref{foursix}), (\ref{fourseven}) 
are connected by a small interval, 
in which
the exponential corrections result in a more 
complicated dependence on
$w$ and $T$.
The above conclusions are verified by the numerical calculation of 
$L^d_n(w,T)$ \cite{trans}.

\setcounter{equation}{0}
\renewcommand{\theequation}{{\bf 5.}\arabic{equation}}

\section*{5. Dimensional reduction}

From this point on we concentrate on $d=4$. 
Our strategy is to solve eq. (\ref{fourthree}) (or its rescaled
form, analogous to eq. (\ref{threetwo}) 
with $d=4$ and $L^4_n(w,T)$ replacing 
$L^4_n(w)$)
for various temperatures, starting the evolution at some very high scale 
equal to the ultraviolet cutoff of the theory $k=\Lx$. 
As the thermal corrections are negligible at this high scale
and no quantum fluctuations have been integrated out yet, 
the initial condition is $U_{\Lx}(\rho,T)=V(\rho)$,
with $V(\rho)$ the classical potential. 
We distinguish 
three regions in the subsequent evolution, as $k$ is lowered from 
$\Lx$ to zero ($\thone$, $\thtwo$ are constants of order one, with 
$\thone < \thtwo$): \\
a)~$ T/k \leq \theta_1$: This is the 
\underline{low-temperature region} where $L^4_{n}(w,T)$ 
are very well approximated by their
zero-temperature value. Also Lorentz invariance is intact and
the $A^0$ component evolves similarly to the other components of 
the gauge field.  
For these reasons we use the zero-temperature 
evolution equation (\ref{twofour}) for $k \geq T/ \theta_1$.
For $T=0$ this region extends all the way to $k=0$.
\\
b)~$\theta_1 <T/k< \theta_2$: 
This is the  \underline{threshold region},
in which $L^4_{n}(w,T)$ do not have a 
simple analytical form. The $A^0$ component starts evolving 
independently of the other components and develops a 
mass term $m^2_{A^0}(k,\rho,T)$. 
\\
c)~$ T/k \geq \theta_2$: 
In the 
\underline{high-temperature region} 
eq. (\ref{fourseven}) gives 
\be 
L^4_{n}(w,T)= 4 \frac{T}{k} L^3_{n}(w).
\label{fiveone} \ee
The three-dimensional character of the effective theory for modes
with $q^2 \ll T^2$ manifests itself in the appearance of the three-dimensional
momentum integrals. It acquires here a precise quantitative meaning.
Arguments from perturbation theory indicate that the 
mass term $m^2_{A^0}(k,\rho,T)$ for the $A^0$ component receives a 
thermal contribution $\sim N_c e^2_R T^2$. As a result this component
soon decouples in this region and its contributions disappear from 
the evolution equations for the potential and the gauge coupling.

Similar regions exist for the evolution of the gauge coupling, even though  
they do not appear explicitly, due to our approximation, which neglects the
threshold functions in the evolution of the gauge coupling. 
It is clear, however, that the evolution of $e^2$ in the low-temperature
region is given by
eq. (\ref{threefive}) 
with $d=4$. 
We approximate the evolution in the threshold region by the same
expression.
In complete analogy with the discussion for the potential, 
the evolution of the gauge coupling in the high-temperature
region is given by 
eq. (\ref{threefive}) 
with $d=4$ and $l^4_{gc}$ replaced by
\be
l^4_{gc}(T)=4 \frac{T}{k} l^3_{gc}. 
\label{fivetwo} \ee
The above expression can be obtained by taking into account the 
threshold functions in the running of $e^2$ \cite{abelian},
considering the non-zero-temperature corrections, and then
setting the masses to zero in the argument of the threshold functions.
In order to be more precise we point out that we are 
assuming that, in the region where eq. (\ref{fivetwo})
applies, the $A^0$ component of the gauge field has
completely decoupled because it has developed a thermal mass. 
This is not strictly true at the beginning of the high-temperature
region, where $k \sim T$, unless the gauge coupling or 
the number of scalar fields is large. 
Notice, however, that even for small coupling the omitted contribution
from the $A^0$ component  
is proportional to the bare coupling and, therefore, small.  
As a result, the error induced 
by this approximation
is expected to be much smaller than 
the errors generated through neglecting the anomalous dimension of 
the scalar field, or by the truncated form of the effective average action.
In this work we do not study the evolution of the
mass term $m^2_{A^0}(k,\rho,T)$. Instead, we rely on the 
perturbative result for the contribution of the $A^0$ field.  
We point out, however, that the explicit study of 
$m^2_{A^0}(k,\rho,T)$ is straightforward in an extended truncation.

In the high-temperature region 
we can define effective three-dimensional parameters
by multiplying with appropriate powers of $T$:
\beq
^3U_k(\rho_3) = &\frac{U_k(\rho,T)}{T}
\nonumber \\
\rho_3 = &\frac{\rho}{T}
\nonumber \\
{e}^2_3(k) = &e^2 (k,T) T,
\label{fivethree} \eeq
and their dimensionless versions
\beq
u_k(\rht) = &\frac{^3U_k(\rho_3)}{k^3}=\frac{U_k(\rho,T)}{k^3 T}
\nonumber \\
\rht = &\frac{\rho_3}{k}=\frac{\rho}{kT}
\nonumber \\
{\et}^2(k) = &\frac{e^2_3  (k)}{k}=\frac{e^2(k,T)T}{k}.
\label{fivefour} \eeq
Making use of eqs. (\ref{fiveone}), (\ref{fivethree}), (\ref{fivefour}),
and assuming the decoupling of the $A^0$ component of the gauge field, 
we can cast
eq. (\ref{fourthree}) with $d=4$ in the form
\beq
\frac{\partial u_k}{\partial t} = & 
~-3 u_k + \rht u'_k
\nonumber \\
&~-v_3 L^3_0 \left( u'_k +2 u''_k \rht \right)
-(2N_c-1) v_3 L^3_0 \left( u'_k \right)
- 2 v_3 L^3_0 \left( 2 \et^2 \rht \right).
\label{fivefive} \eeq
This is nothing but eq. (\ref{threetwo})
with $d=3$. We started with the evolution equation for the
four-dimensional theory at non-zero temperature, and, in the 
high-temperature region, derived the evolution equation
for an effective three-dimensional theory at zero temperature.
Similarly, for the gauge coupling we find
\be
\frac{d \et^2}{dt} = 
-\et^2 + \frac{4}{3} v_3 N_c \et^4 l^3_{gc}.
\label{fivesix} \ee
We have managed to reduce the problem to the
set of eqs. (\ref{fivethree})--(\ref{fivesix}). These equations
have to be supplemented with appropriate initial conditions 
for the form of the potential and the value of the gauge
coupling at the beginning of the high-temperature region
$k=T/\theta_2$. 
Before addressing this point, we note that our results do not
imply that the four-dimensional
non-zero-temperature problem can always be
reduced to the three-dimensional one.
The evolution equation (\ref{fivefive}) involves threshold functions,
which could switch off the evolution if the running masses of the
various modes are large in the high-temperature region.
Then the form of the potential at $k=0$ would be determined 
to a large extent by the initial conditions for the running in
the high-temperature region. These initial conditions
are determined by the evolution in 
the low-temperature and intermediate regions and depend strongly
on the structure of the four-dimensional theory.
The dynamics of the three-dimensional theory becomes dominant 
in the case of a second-order 
or weakly first-order transition near the critical 
temperature, where
the running masses stay small, or if there is a strongly-coupled
three-dimensional regime.

\setcounter{equation}{0}
\renewcommand{\theequation}{{\bf 6.}\arabic{equation}}

\section*{6. The initial conditions for the
three-dimensional evolution}

We now turn to the question of determining the initial
conditions for the evolution in the effective three-dimensional
region $k \leq T/ \theta_2$. A straigtforward integration of the
evolution equations would relate the potential and coupling at
$k = T/\theta_2$ to the classical potential and bare coupling at
$k=\Lx$. However, a more physically transparent picture is 
obtained if we trade the classical parameters for those of the 
renormalized theory at zero temperature. An important simplification
is provided by the fact that the evolution between 
$k=\Lx$ and $k = T/\theta_1$ is identical for the zero and non-zero
temperature theories. 
Moreover, at zero temperature
the running of the various couplings
between $k = T/\theta_1$ and $k=0$ is logarithmic 
and can be neglected for sufficiently small values of the couplings. 
Similarly, threshold effects in the evolution equation for the 
potential can be neglected, 
as the relevant mass terms are proportional to the couplings.
This again is a minor approximation compared to the omission of terms 
in the truncated effective action. Relaxing this approximation is possible,
through the numerical integration of the evolution \cite{trans}. 
However, in order 
to derive analytical expressions and 
not complicate our analysis (without significant improvement in accuracy)
we neglect this logarithmic running at this stage.
We shall later account for most of the resulting error by comparison
with the perturbative result (see the end of this section). 

Let us consider a quartic, renormalized, zero-temperature potential.
Higher terms in the potential  are set equal to zero, consistently 
with our approximation of no running for the couplings.
We parametrize it as
\footnote{The expression (\ref{sixone}) corresponds to the
non-convex outer part of the potential. The inner part is 
totally flat and irrelevant for the determination of the initial
conditions for the three-dimensional evolution.
We have also assumed that we are away from the parameter region where 
radiative symmetry breaking \cite{colwein} 
generates a two-minimum structure in four dimensions. 
This typically happens for
the range $\lx_R/e^4_R \simeq 1/4\pi^2$, which is not considered here.
For a study of the Coleman-Weinberg phase transition in four dimensions, in
the context of the effective average action, see ref. \cite{abelian4d}. }
\be
U'(\rho) \equiv U'_0(\rho) = \lx_R \left( \rho - \rho_{OR} \right),
\label{sixone} \ee
and we also define the renormalized gauge coupling
\be
e^2_R \equiv e^2(0).
\label{sixtwo} \ee
The zero-temperature evolution equation
(\ref{threethree}) with $d=4$ can be integrated between 
$k=0$ and $k = T/\theta_1$, if the running of the couplings and the threshold
effects are neglected. We obtain for the $\rho$ derivative of the potential
\be
U'_{T/\thone}(\rho,T) = \lx_R 
\left\lbrace 
\rho - \rho_{0R} - 
\left[ (2 N_c + 2) + 6  \frac{e^2_R}{\lx_R} \right]
v_4 \left( \frac{T}{\thone} \right)^2 
\right\rbrace.
\label{sixthree} \ee
We observe that the effect of the evolution is a
shift of the location of the minimum proportional to the square of
the scale $T/\thone$ (a quadratic renormalization of the minimum).
We can make the same approximations
in order to integrate the non-zero-temperature evolution 
in the intermediate region between
$k = T/\theta_1$ and $k = T/\theta_2$.
However, information on the behaviour of the mass 
$m^2_{A^0}(k,\rho,T)$ for the $A^0$ component is required in this
region. For this reason we separate the contribution coming from
the $A^0$ field in order to obtain
\beq
U'_{T/\thtwo}(\rho,T) =~&U'_{T/\thone}(\rho,T) 
+ \lx_R 
\left[ (2 N_c + 2) + 4 \frac{e^2_R}{\lx_R} \right]
2 v_4 T^2 I + C^{\rm int}_{A^0} 
\nonumber \\
=~&\lx_R \left\lbrace 
\rho - \rho_{0R} - 
\left[ (2 N_c + 2) + 4 \frac{e^2_R}{\lx_R} \right]
2 v_4 T^2 \left( \frac{1}{2 \thone^2} - I \right) 
\right\rbrace 
\nonumber \\
&-2 v_4 e^2_R
\left( \frac{T}{\thone} \right)^2 
+ C^{\rm int}_{A^0}.
\label{sixfour} \eeq
Here 
\be
I = \int_{1/\thtwo}^{1/\thone} dy~y t^4_1 \left( 0, \frac{1}{y} \right),
\label{sixfive} \ee
with
\be
t^4_1 \left( 0,\frac{T}{k} \right) =
\frac{L^4_1 \left( 0,\frac{T}{k} \right)}{L^4_1(0,0)}
= 2 \sqrt{\pi} \frac{T}{k} \sum_{n=-\infty}^{\infty}
\exp \left\lbrace - 4 \pi^2 n^2 \left( \frac{T}{k} \right)^2 \right\rbrace.
\label{sixsix} \ee
For sufficiently small $\thone$ and large $\thtwo$
\be
I - \frac{1}{2 \thone^2} + 2 \sqrt{\pi} \frac{1}{\thtwo} 
\simeq \int_{0}^{\infty} dy~y \left[ 
t^4_1 \left( 0, \frac{1}{y} \right) -1 \right]
= \frac{2 \pi^2}{3}.
\label{sixseven} \ee 
The last two terms in the r.h.s. of eq. (\ref{sixfour})
are parts of the thermal correction to the effective potential 
coming from the fluctuations of the $A^0$ component. 
As we explained in the previous section after eq. (\ref{fivetwo}),
we expect another
small contribution coming from the evolution in the 
high-temperature region, until the complete decoupling of the
$A^0$ field due to its thermal mass. 
Perturbation theory should give a good approximation for the
total contribution, as no infrared problems are associated with
the $A^0$ fluctuations, due to their non-zero thermal mass. 
For this reason we omit the 
$A^0$ field completely in the high-temperature region and 
include the perturbative result for its 
contribution in the initial conditions at $k=T/\thtwo$.
This leads to 
\beq
U'_{T/\thtwo} (\rho,T)
=~&\lx_R \left\lbrace 
\rho - \rho_{0R} - 
\left[ (2 N_c + 2) + 4 \frac{e^2_R}{\lx_R} \right]
2 v_4 T^2 \left( \frac{2 \sqrt{\pi}}{\thtwo} - \frac{2 \pi^2}{3} \right) 
\right\rbrace
\nonumber \\
&+U'_{A^0} (\rho,T),
\label{sixeight} \eeq
with 
\beq
U'_{A^0}(\rho,T) =~&
\frac{d}{d \rho} \left\lbrace 
\frac{T^4}{2 \pi^2} \int_0^{\infty} dx~x^2 \ln \left( 
1- \exp \left[ - \sqrt{ x^2 + \frac{m^2_L}{T^2} } \right]
\right) \right\rbrace
\nonumber \\
=~& \frac{d}{d \rho} \left( \frac{1}{24} m^2_L T^2 
-\frac{1}{12 \pi} m^3_L T + ... \right),
\label{sixnine} \eeq
and 
\be
m^2_L = 2 e^2_R \rho + \frac{N_c}{3} e^2_R T^2.
\label{sixten} \ee
Equation (\ref{sixnine}) is the perturbative one-loop contribution
of the $A^0$ component, whose total mass term, including the 
thermal Debye part, is given by eq. (\ref{sixten})
\cite{buchabel,hebecker,shapred}.
The initial condition for the 
gauge coupling is given by 
\be
e^2(T/\thtwo,T) = e^2_R.
\label{sixeleven} \ee
This completes the first part of our study. We have 
reduced the problem to the three-dimensional one for
the couplings defined in eqs. (\ref{fivethree}), (\ref{fivefour}).
The evolution equations are given by eqs. (\ref{fivefive}), (\ref{fivesix}).
For the numerical part we find it more convenient to work with the
evolution equation for $u'_k$, given by eq. (\ref{threethree}) with $d=3$.
The initial conditions (at $k=T/\thtwo$)
for the evolution in the three-dimensional
regime are given by eqs. (\ref{sixeight})--(\ref{sixeleven}).
Large values of $\theta_2$ minimize the error introduced by 
the approximate treatment of the $A^0$ component. However, $\theta_2$
must be taken sufficiently small, so that the full three-dimensional evolution 
is properly taken into account.
We find that the optimal value is $\theta_2= 1$. 

Before proceeding with our analysis, it is 
instructive and reassuring 
to check the validity of our approximations through 
a comparison with known results from perturbation theory.
If we integrate the evolution equation (\ref{threethree})
with $d=3$, neglecting the running of couplings and threshold effects,
we find for the temperature-dependent potential
\beq
U'(\rho,T) \equiv U'_{0}(\rho,T) =~&U'_{T/\thtwo}(\rho,T) 
+ \lx_R 
\left[ (2 N_c + 2) + 4 \frac{e^2_R}{\lx_R} \right]
v_3 \sqrt{\pi} T^2 \frac{1}{\thtwo}
\nonumber \\
=~&\lx_R \left\lbrace
\rho - \rho_{0R} + 
\left[ (2 N_c + 2) + 6 \frac{e^2_R}{\lx_R} \right]
\frac{T^2}{24}
\right\rbrace,
\label{sixtwelve} \eeq
where we have kept only the leading term in the expansion of the
$A^0$ contribution of eq. (\ref{sixnine}). This predicts a critical
temperature
\be
T^2_{cr} = 
\frac{24 \rho_{0R}}{(2 N_c + 2) + 6 \frac{e^2_R}{\lx_R}},
\label{sixthirteen} \ee
in agreement with the lowest-order perturbative result
\cite{buchabel,hebecker}.
Corrections to the above expression arise through the 
inclusion of the running of couplings and threshold effects in the
evolution. 
Another check can be made by noticing that, if 
the contributions from the scalar fields and the running of 
couplings are neglected, eq. (\ref{fourthree}) can be easily integrated
in order to reproduce
the one-loop perturbative contribution coming from the gauge field.
In our approach we have reformulated 
eq. (\ref{fourthree}) in terms of the 
three-dimensional evolution with an appropriate boundary condition. 
As a check we present in fig. 1 
the results of the numerical integration
of eq. (\ref{fivefive}), in which we have omitted
the scalar contributions and the running of $e^2$. 
We have used 
$N_c=1$, $\lx_R=0.02$, $e^2_R=0.09$,
$T^2/\rho_{0R} \simeq 0.953$.
Line (a) is the perturbative one-loop result coming from gauge-field
contributions.
Line (b) results from the numerical integration of the 
evolution equation (\ref{fivefive}),
if we neglect the first and second terms in the second line, which
come from the scalar modes.
The observed discrepancy can be attributed to
the omission of the threshold corrections in the four-dimensional
running, which determines the initial condition of eq. (\ref{sixeight}).
The biggest part of the discrepancy is due to an imprecise
value for the renormalization of the location of the minimum, for
which we have used
\be
\Delta\rho_0(k=T/\theta_2,T) = 8 \frac{e^2_R}{\lx_R}
v_4 T^2 \left( \frac{2 \sqrt{\pi}}{\thtwo} - \frac{2 \pi^2}{3} \right)
= -0.32981 \rhz.
\label{sixfourteen} \ee
Due to the threshold effects 
this value receives corrections proportional to powers
of the couplings $\lx_R$, $e^2_R$. For the cases we study in the
following, the largest corrections are proportional to $e^4_R$.
Line (c) is the result of the numerical integration if the minimum of the
above potential is shifted by an additional amount 
$\delta\rho_0(k=T/\theta_2,T)
=-0.00082 \rhz$ for the same critical temperature.
We observe very good agreement. 
We can alternatively view the need for 
the additional shift of the minimum as a correction to the value of
the critical temperature that we have used. 
This means that our value for the critical temperature 
cannot be trusted to better than 0.3\%. However, approximations associated
with the truncated form of the effective average action are expected to 
generate even larger errors. In any case, 
in order to eliminate this particular source of uncertainty, we 
repeat the empirical procedure that we have used above for every model
we study in the following.
Through comparison with the perturbative result, we determine an
empirical value $\delta\rho_0(T/\theta_2)$, which we incorporate 
in the initial condition of eq. (\ref{sixeight}).

\setcounter{equation}{0}
\renewcommand{\theequation}{{\bf 7.}\arabic{equation}}

\section*{7. The phase diagram and the second-order phase transitions}

In the previous sections we developed the formalism that we
need for the study of the phase transitions in the Abelian Higgs
model. We reduced the problem of calculating the 
temperature-dependent effective potential to the question of 
integrating the evolution equations (\ref{fivefive})
(or equivalently eq. (\ref{threethree}) with $d=3$) and (\ref{fivesix}),
for the effective three-dimensional parameters (effective average 
potential and running gauge coupling)
defined in eqs. (\ref{fivethree}), (\ref{fivefour}). 
The evolution starts at the scale $k=T/\thtwo$ (we use $\thtwo=1$),
with the initial conditions given by 
eqs. (\ref{sixeight})--(\ref{sixeleven}) in terms of the renormalized
parameters of the zero-temperature theory.
The temperature-dependent potential is obtained in the limit
$k \rightarrow 0$. 
The initial condition for the potential incorporates the contributions
from the part of the evolution that integrates out fluctuations with
four-dimensional character. The remaining part is purely three-dimensional
and the universal results of our study (such as the existence of fixed points)
are characteristic of the three-dimensional theory. For this reason 
they are relevant for all theories that belong to the
same universality class. For example, they can be applied to the 
theory of the superconductor phase transition. 
Contact with perturbation theory can be made 
if one neglects the contributions 
to eq. (\ref{fivefive}) that come 
from the scalar modes (the first and second terms in the second line)
and the evolution of $e^2$ (by neglecting the second term in 
the r.h.s. of eq. (\ref{fivesix})).
The integration reproduces the part of the 
one-loop result that comes from the fluctuations of the gauge field. 
If the scalar field contributions are included, but the gauge-coupling
running is still neglected, agreement with perturbation theory is not
possible. The reason is that perturbation theory at a finite order
cannot correctly describe the relevant
infrared physics associated with massless (or almost massless)
scalar modes. Resummations of infinite subclasses of 
perturbative diagrams through gap equations \cite{doljac,buchgap,owe,largen}
may ameliorate the situation, 
but even they do not predict correct values
for quantities such as critical exponents. We shall return to this point
in the following. The inclusion  of the running of the gauge coupling
generates further deviations from the perturbative results.

We start by discussing the phase diagram for the Abelian Higgs
model with $N_c$ complex scalars.
The expected qualitative behaviour can be inferred by considering the
evolution equations (\ref{fivesix}) and (\ref{threeeight}) (with $d=3$),
for the gauge and quartic couplings. We point out that 
eq. (\ref{threeeight})
assumes a quartic form for the potential and no threshold effects, 
and for this reason 
the predicted behaviour is not trustworthy. However, it is instructive to
consider this simplified system of equations first and 
perform the complete analysis later. 
The flows of couplings are depicted in fig. 2 for a theory 
with $N_c=250$.
They exist on a critical surface, with the minimum of the potential 
(or the mass term)
as the relevant parameter. One may consider the
flows as being unstable perpendicularly to the plane of fig. 2. 
Moving above or below the critical surface
corresponds to a growing or diminishing 
value for the minimum, 
and the respective deviations
lead to the phase with symmetry breaking or the symmetric one 
\cite{trans,indices}. 
We observe the existence of three fixed points on the critical surface:\\
a) The Wilson-Fisher fixed point (WF) has $\et^2=0$
and corresponds to the pure scalar theory. 
It is unstable in the $\et^2$ direction. \\
b) The Abelian fixed point (A) is
the most stable one and is expected to be relevant very close to the
critical temperature of the second-order phase transitions. \\
c) The tricritical point (T) separates the region 
of second-order phase transitions from the region of first-order ones.
The latter are expected when the quartic coupling $\lt$ turns negative,
indicating the appearance of an instability \cite{amit,twoscalar}.\\
The above phase diagram ceases to exist below 
$\left( N_c \right)_{cr} = 222$, where the Abelian and tricritical 
points disappear and all the flows end up in the region of 
first-order phase transitions. This value of $\left( N_c \right)_{cr}$
is of the same order of magnitude as the 
prediction from the $\ex$-expansion \cite{ginsparg}. 

We now replace the evolution equation (\ref{threeeight}) with 
the partial differential equation
(\ref{fivefive})
for the full potential. This equation automatically 
takes into account the running of all the higher couplings 
and the threshold effects. In practice, we find more efficient to 
integrate eq. (\ref{threethree}) with $d=3$.
Two algorithms for the numerical integration 
have been presented in detail in ref. \cite{num}.
The comparison of the two methods 
provides a good check for possible systematic numerical 
errors. The two algorithms give results which 
agree at the 0.3\% level. We expect 
the numerical solution to be an approximation of the solution
of the partial differential equation (\ref{threethree}) with the same 
level of accuracy. 
In fig. 3 we present the results of the numerical integration
of eq. (\ref{threethree}) for $d=3$ and $N_c=5$.
We consider a renormalized, zero-temperature potential 
given by eq. (\ref{sixone}) with 
$\lx_R=0.5$. 
The renormalized gauge coupling at zero temperature
is chosen as $e^2_R=10^{-6} \times e^2_{A}$, where $e^2_{A}$ is its
value at the Abelian fixed point. 
The temperature is very close to the critical one $\tcr^2/\rhz \simeq 2.10$.
(We use the value of the zero-temperature minimum in order to renormalize
all dimensionful quantities.)
The function $u'_k(\rht)$ is plotted
for various values of $k$, starting from the beginning of the 
high-temperature region $k=T/\thtwo$ (dotted lines). As $k$ is lowered 
the system approaches first the Wilson-Fisher fixed point, and the
evolution slows down around the scale-invariant solution corresponding
to this fixed point (solid lines). 
During this part of the evolution the gauge coupling
$\et^2$ stays small and the theory is dominated by the scalar modes.
Subsequently the gauge coupling becomes significant and acts as a relevant
perturbation, which 
causes the system 
to leave the Wilson-Fisher fixed point and approach the   
Abelian one. Eventually $u'_k(\rht)$ moves away from the Abelian fixed point
too, and the
final running leads to the phase with 
spontaneous symmetry breaking (dashed lines).
The relevant perturbation for this last deviation is the value of 
the minimum of the
potential at the beginning of the running 
or, equivalently, the temperature. 
With sufficient fine-tuning of the temperature 
the system can spend a long ``time'' 
$t = \ln \left(k/\rhz^{1/2} \right)$ 
in the vicinity of the fixed points. During this
``time'', the minimum $\kx$ of the potential $u_k(\rht)$ 
varies very little, 
while the minimum of $U_k(\rho,T)$ evolves towards zero according to 
\be
\rho_0(k,T) = k \kx.
\label{sevenone} \ee
The longer $u'_k(\rht)$ stays near the fixed points, the 
smaller the resulting value of $\rho_0(k,T)$ when the 
system deviates from it. 
As this value determines the mass scale for the
renormalized theory at $k=0$, the 
fixed-point solutions govern the behaviour of the system 
very close to the phase transition, where the characteristic 
mass scale goes to zero. 
As a result the value of the temperature that keeps the 
system close to the fixed points for an arbitrarily long ``time''
is the critical one. 
Also, the minimum of the potential and the characteristic mass scale 
go continuously to zero as the critical temperature is approached. 
This demonstrates the presence of second-order phase transitions when
the behaviour of the system is governed by the fixed points. 
The approach to the fixed-point solutions and the deviation
from them can also be seen in fig. 4.
We plot the evolution of the minimum $\kx(k)$ of $u_k(\rht)$,
the coupling $\lt(k)= u''_k(\kx)$, 
and the gauge coupling $\et^2(k)$.  
All higher couplings, related to higher derivatives of $u_k$,
can be easily obtained from the solution of fig. 3. 
For the solution depicted in
figs. 3 and 4
the temperature is slightly below $\tcr$. The potential
$u_k(\rht)$ moves away from the Abelian fixed point  
and its minimum
$\kx(k)$ grows in such a way that
$\rho_0(k,T)$ approaches a constant value
for $k \rightarrow 0$.
Eventually the theory settles down in the 
phase with spontaneous symmetry breaking with a renormalized value
$\rho_{0R}(T)$ for the minimum of $U(\rho,T)$.
If the temperature is larger than $\tcr$,
the minimum $\kx(k)$
runs to zero and $u'_k(0)$ becomes positive and 
subsequently increases. 
The system settles down in the symmetric phase
with a positive constant renormalized mass term 
$m^2_R(T) = k^2 u'_k(0) $ as $k \rightarrow 0$. 
We point out that larger deviations from $\tcr$
cause the system to deviate towards the phase with spontaneous
symmetry breaking or the symmetric one earlier in the evolution.
As a result, the relative influence 
of the various fixed points on the renormalized theory
depends sensitively on the precise value of the temperature. 

Another important property of the ``near-critical''
trajectories, which spend a long ``time'' $t$ near
the fixed points, is that they become insensitive
to the details of the zero-temperature theory, which determine the
initial conditions for the evolution in the three-dimensional
region.
After $u'_k(\rht)$ has
evolved away from its scale-invariant form near a fixed point,
its shape is independent of the choice of $\lx_R$ and $e^2_R$. 
This property gives rise to the universal critical behaviour
near second-order phase transitions. 
As the Abelian fixed point is the most attractive one,
it is expected to determine the characteristics of the phase transition
very close to the critical temperature. 
The most typical quantities parametrizing the behaviour near the
transition are the critical exponents, which are universal quantities depending
only on the dimensionality of the system and its internal
symmetries. We concentrate on
the exponent $\nu$, which parametrizes the behaviour of the
renormalized mass in the critical region. Another critical 
exponent is the anomalous dimension $\eta$ of the scalar field, which
we have approximately taken to be zero. The other exponents
are not independent quantities, but
can be determined from $\eta$ and $\nu$ through universal 
scaling laws. We define the exponent $\nu$ through
the renormalized mass term in the symmetric phase 
$m^2_R(T) = k^2 u'_k(0) $ for $k \rightarrow 0$. 
The behaviour of $m^2_R(T)$ in the critical region 
depends only on the distance from the phase transition, which
can be expressed in terms of the difference of the temperature
$T$ from the critical temperature $T_{cr}$, for which
the renormalized theory has exactly $m^2_R(\tcr) =0$. 
The exponent $\nu$ is defined through the relation
\be
m^2_R(T) \sim \left( T^2-\tcr^2 \right)^{2 \nu}.
\label{seventwo} \ee
For a determination of $\nu$ from our results we calculate
$m^2_R(T)$ for various values of $T^2-\tcr^2$.
We subsequently plot $\ln(m^2_R)$ as a function
of $\ln \left( T^2-\tcr^2 \right)$ and infer an ``effective'' exponent 
$\nu$ from the slope.
In fig. 5 we plot this ``effective'' exponent for the model of fig. 3 
as the critical temperature is approached. 
Each point along the curve corresponds
to trajectories similar to those of figs. 3 and 4, which approach 
and subsequently deviate from the fixed points at various 
``times'' $t=\ln \left( k/\rhz^{1/2} \right)$. 
For the model of fig. 3,
the initial conditions (determined by the zero-temperature theory)
place the system initially close to the Wilson-Fisher fixed point.
As the critical temperature is approached, the system stays longer
near the fixed points before deviating towards the symmetric phase.
As the Abelian fixed point is the most attractive one, it dominates 
the largest part of the evolution for very small $T-\tcr$. 
This happens because the gauge coupling has enough ``time'' to grow towards
the attractive fixed point of eq. (\ref{fivesix}).
This behaviour is reflected on the ``effective'' exponent $\nu$.
It first approaches a value typical of the Wilson-Fisher fixed point,
but very close to $\tcr$ it settles down at 
a value characteristic of the Abelian fixed point. 
The part of the curve between the Wilson-Fisher and the Abelian fixed points
is typical of a crossover curve \cite{twoscalar,crossover}.
Values of the critical exponent $\nu$ for the Wilson-Fisher fixed point,
for various $N_c$, calculated through the method of the effective 
average action,
have been presented in refs. \cite{indices,num,eos}. In table 1 
we give values characteristic of the Abelian fixed point. They are consistent
with the results of ref. \cite{supercond}, if one takes into account
the difference in the approximations employed. (We have neglected the
anomalous dimension of the scalar field and used a general form for the
potential, whereas in ref. \cite{supercond} the anomalous dimension
is taken into account, but the potential is approximated by a polynomial.)
However, in contrast to ref. \cite{supercond}, 
we find that the tricritical point and the Abelian fixed point
disappear for $N_c < 5$. The
two approaches are compared at the end of the section.

In order to complete the phase diagram, we 
must investigate the region that leads to first-order phase transitions.
In the simplified picture of fig. 2 a tricritical point (T) separates the
regions of first-order and second-order phase transitions. The form of the
full potential at this point is presented in fig. 6, along with its 
form at the Wilson-Fisher (WF) and the Abelian (A) fixed points.
At the tricritical point the potential develops a new unstable minimum
at the origin ($u_k'(0)$ is positive). 
Another characteristic point is the inflection point (I),
where the minimum of the potential away from the origin disappears and 
the origin remains the only minimum.
Various patterns of evolution can produce interpolations between
the different forms of the potential presented in fig. 6. A typical
example was given in fig. 3. For other values of $\lx_R$ and  
$e^2_R$
the potential may evolve first towards the tricritical point,
and subsequently be attracted to the Abelian fixed point, before deviating
towards the phase with spontaneous symmetry breaking or the symmetric one.
An unstable minimum appears for part of the evolution, but eventually
disappears. 
Flows of this type again give second-order phase transitions.
For the same $e^2_R$ and 
smaller $\lx_R$ the flow leads from the tricritical point to the inflection
one. Along the way the absolute minimum of the potential shifts from
the minimum away from the origin to the origin. This typical behaviour
characterizes first-order transitions. It has been observed in the past
in the 
context of the effective average action, using cruder approximations
\cite{twoscalar,twoscalar4d,abelian4d}. 
We study the first-order phase transitions in detail in the next
section.

We conclude this section by comparing our results with those of other 
approaches. Perturbation theory cannot reproduce the behaviour associated
with fixed points. It cannot capture the dynamics of massless (or almost
massless) scalar modes. As a result, no part of the rich structure that
we presented in this section is visible to it. 
The leading result from the 
$\ex$-expansion is in agreement with the qualitative 
structure of the phase diagram of fig. 2 for a 
large number of scalar fields
$N_c = {\cal{O}}$(100) \cite{ginsparg}. 
In contrast, we have found that the tricritical point and the 
Abelian fixed point persist down to $N_c=5$.
We believe that our conclusion gives a more reliable 
estimate of the order of magnitude of 
$\left( N_c \right)_{cr}$, below which
the phase diagram of fig. 2 ceases to exist, because
our analysis was carried out directly in three dimensions. 
A very important question is whether the 
fixed-point structure that we discussed exists also for $N_c=1$.
This would raise the possibility of a second-order superconducting
phase transition \cite{secsup}. The analysis of ref. \cite{supercond}
supports this possibility through the study of the same renormalization-group 
equation for the effective average action. The difference lies in
the approximations involved. We have solved numerically the
evolution equation for the full potential without using a
Taylor expansion around a minimum, which was done in ref. 
\cite{supercond}. This has permitted us 
to investigate the appearance of a second minimum without resorting to
extrapolations. We have neglected effects 
that are expected to be small, such as the 
anomalous dimension of the scalar field and threshold effects in the
running of the gauge coupling. 
However, a major source of uncertainty in both approaches
is related to the assumed truncated form of the
effective average action. Effects coming from the $\rho$ dependence of
the gauge coupling, higher derivative terms, and 
the role of the background gauge field $\Ab$ have not been taken into account.
An extended truncation is needed in order to estimate their magnitude. 
At the present stage we can conclude that 
the phase diagram of fig. 2 survives down to a critical number of scalar
fields 
$\left( N_c \right)_{cr} = {\cal{O}}(1)$.
This supports the possibility of a second-order superconductor 
phase transition, but does not guarantee it.

\setcounter{equation}{0}
\renewcommand{\theequation}{{\bf 8.}\arabic{equation}}

\section*{8. The first-order phase transitions}

We now turn to the detailed study of the region of 
first-order phase transitions. We concentrate on the case $N_c=1$ because
it is the most relevant for 
physical applications, such as the non-zero-temperature behaviour
of superconductors. 
The analysis is very similar for other values of $N_c$. 
The expected phase diagram can again be deduced from the 
set of equations 
(\ref{fivesix}) and (\ref{threeeight}) (with $d=3$),
for the gauge and quartic couplings. It is depicted in fig. 7.
We observe that the Abelian fixed point and the tricritical point
have disappeared and all the flows end up in the region of 
first-order phase transitions. 
We now replace the evolution equation (\ref{threeeight}) with 
the partial differential equation
(\ref{fivefive})
for the full potential. 
This takes into account the running of all the higher couplings 
and the threshold effects. 
In fig. 8 we present the results of the numerical integration
of the evolution equation for the potential for $N_c=1$.
We consider a renormalized, zero-temperature potential 
given by eq. (\ref{sixone}) with 
$\lx_R=0.5$. 
The renormalized gauge coupling at zero temperature
is chosen as $e^2_R=10^{-7} \times e^2_{A}$, where $e^2_{A}$ is its
value at the Abelian fixed point. 
The temperature is very close to the critical one $\tcr^2/\rhz \simeq 6.41$.
The function $u'_k(\rht)$ is plotted
for various values of $k$, starting from the beginning of the 
high-temperature region $k=T/\thtwo$ (dotted lines). As $k$ is lowered 
the system approaches first the Wilson-Fisher fixed point, and the
evolution slows down around the scale-invariant solution corresponding
to this fixed point (solid lines). 
During this part of the evolution the gauge coupling
$\et^2$ stays small and the theory is dominated by the scalar modes.
Subsequently the gauge coupling becomes significant and acts as a relevant
perturbation, which 
causes the system 
to leave the Wilson-Fisher fixed point.
In contrast to fig. 3 no other fixed points exist which could attract the
flow of the potential. 
At some point during the evolution 
(dashed lines)
a new minimum appears at the origin ($u'_k(0)$ turns positive).
Subsequently it becomes 
deeper than the initial minimum,
and eventually the only minimum of the potential.
The behaviour at the later stages of the evolution 
is characteristic of a first-order phase transition.
By slightly reducing the temperature we could arrange for the potential
$U_k(\rho,T)$ to end up with two minima of equal depth
for $k \rightarrow 0$. We explicitly demonstrate how this happens 
in the rest of this section. In fig. 9 we display the evolution of the
original minimum $\kx$ of $u_k(\rht)$, the quartic coupling 
$\lt(k)=u''_k(\kx)$ and the gauge coupling $\et^2(k)$.
We observe how $\lt(k)$ goes to zero at the point where
the original minimum disappears. 

The Wilson-Fisher fixed point is important only for very small 
values of the zero-temperature coupling $e^2_R$. 
We are interested in larger values 
of $e^2_R$ (comparable to $\lx_R$) in order to develop some 
intuition relevant for the electroweak phase transition.  
For this range of $e^2_R$ the scale-invariant solution of the
evolution equation for the potential
is never approached. Similarly, the 
question of the persistence of the Abelian fixed point and the 
tricritical point down to $N_c=1$ (see end of last section)
is not relevant. The characteristic evolution of the potential
$U_k(\rho,T)$ as $k$ is lowered is depicted in fig. 10. 
In order to produce fig. 10, we have solved numerically 
the evolution equation for the derivative of 
the rescaled potential $u'_k(\rht)$ as before,
and deduced $U_k(\rho,T)$ through eqs. (\ref{fivethree}), (\ref{fivefour}). 
The zero-temperature theory is 
given by eq. (\ref{sixone}) with 
$\lx_R=0.02$, and $e^2_R=0.09$. The temperature is 
$T^2/\rhz \simeq 0.841$.
Initially the potential has only one minimum 
away from the origin, which receives 
a renormalization proportional to 
the running scale $k$ during the evolution, and moves closer to the origin.
At some point a new minimum appears at the origin. 
It is induced by 
the integration of thermal fluctuations, through the
generalization of the Coleman-Weinberg mechanism.
The evolution slows down at the later stages and the 
potential converges towards a non-convex profile with two
minima of equal depth. Around the minima
the scale $k$ becomes smaller
than the mass of the various massive modes,
and this induces their decoupling. 
We have stopped the evolution at a non-zero $k_{b}$, for which the
shape of the potential is most stable. The presence of the
non-convex part is explained by this non-zero value of $k$.
We have not yet integrated out all the thermal fluctuations, 
which should render the effective potential convex. More specifically,
we have not taken into account fluctuations that interpolate 
between the two minima of fig. 10 \cite{convex}. They are the ones
that trigger the thermal tunnelling and drive the first-order 
phase transition. If we continue the evolution all the way
to $k=0$, these interpolating configurations will be  
gradually integrated out.
As a result, the height of the barrier will start getting smaller, until
the region of the potential between the two minima becomes flat.

For the model of fig. 10 the behaviour of the potential
suggests the separation of the study of the first-order phase transition
into two parts. First, the evolution equation is integrated down to a given
scale $k_b$, where the shape of the potential becomes approximately stable
around a non-convex
form. The scale $k_b$ can be identified with the coarse-graining scale,
at which the free energy is defined in all studies of statistical systems. 
The effective average action at this scale plays the role of the
coarse-grained free energy. Our formalism provides the 
necessary framework for the derivation of the coarse-grained free energy
from the bare action. This should be contrasted with the perturbative 
approach, in which the notion of coarse graining is absent. 
The second part of the problem concerns the 
dynamics of the first-order phase transition 
in terms of the well-defined potential. 
In the context of the nucleation picture \cite{langer},
semiclassical techniques
can be employed in order to estimate the nucleation probability, 
based on a dominant configuration that  
is usually referred to as the critical bubble \cite{lindebub}. 
The characteristics of the critical bubble can be deduced from the
form of the potential and we present them in the following.

In fig. 11 we compare various approximations of the 
form of the potential at
the respective critical temperatures of the first-order phase transitions for
the model of fig. 10. 
Line (a) is the perturbative one-loop result if the 
contribution of the scalar fluctuations is neglected. The integration
of the evolution equation, if the terms involving scalar modes and
the running of the gauge coupling are neglected, reproduces this
result. 
Line (b) is the perturbative one-loop result when 
the scalar fluctuations are taken into account. We have used 
expressions analogous to eq. (\ref{sixnine}), with thermally corrected masses
given by 
\beq
m^2_G = &\lx_R \rho + 
\left( \frac{1}{4}e^2_R + \frac{2 N_c +2}{24}\lx_R \right)
\left(T^2 - \frac{24 \rho_{0R}}{(2 N_c + 2) + 6 \frac{e^2_R}{\lx_R}}
\right)
\nonumber \\
m^2_R = &3 \lx_R \rho + 
\left( \frac{1}{4}e^2_R + \frac{2 N_c +2}{24}\lx_R \right)
\left(T^2 - \frac{24 \rho_{0R}}{(2 N_c + 2) + 6 \frac{e^2_R}{\lx_R}}
\right)
\label{eightone} \eeq
for the Goldstone and radial modes, respectively.
Due to the
smallness of the quartic coupling $\lx_R$,
the effect of the scalar fluctuations is minor.
Line (c) results from the integration of the full evolution equation 
for the potential down to the scale $k_b$,
if the running of $e^2$ is neglected. 
We observe that the predicted strength of the first-order phase
transition is much smaller than what is expected from one-loop perturbation
theory (line (b)). This is due to the inability of perturbation theory
to properly describe
the infrared dynamics of massless (or almost massless)
scalar fluctuations. The same problem has also been encountered  
in the study of the phase transition for the pure scalar theory. 
Perturbation theory at any finite order predicts a first-order
phase transition for this system \cite{scalpert}. The resummation of
an infinite subclass of perturbative diagrams through gap equations 
\cite{doljac,buchgap,owe,largen} improves the situation, but only
the use of the renormalization group \cite{trans}
gives the full picture of a second-order phase transition characterized
by critical exponents. 
Line (d) results from the integration of the evolution equation 
for the potential
with the running of $e^2$ included.
The theory is infrared-free and the running gauge coupling gets 
smaller when $k$ is reduced. As the first-order phase
transition is triggered by the gauge field fluctuations, its
strength is reduced because the ``effective'' gauge coupling 
is smaller than $e^2_R$. We should also mention that
a reduction of the strength of the phase transition 
is observed in higher orders of perturbation theory, where the
running of the couplings is partly accounted for \cite{hebecker}.

The characteristics of the phase transition can be deduced from
the form of the potential. The most natural scale for the determination
of the various parameters is the  
coarse-graining scale $k_b$. This is convenient, because
the masses of various fields are often renormalized to zero for 
$k \rightarrow 0$, due to approximations or the presence of 
massless modes in the evolution. 
For example, the running of the gauge coupling (which determines the mass
of the gauge field) never stops, due to the omission of threshold effects
in the evolution equation (\ref{fivesix}). 
Also the running of the quartic coupling at the minimum away from zero
(which determines the mass of the radial scalar mode)
never stops. This is 
due to the presence of Goldstone modes at this point, which
never decouple. 
In table 2 we summarize the characteristics of the phase transition
for the model of fig. 10.
The critical temperature and the discontinuity 
in the order parameter are given in the first three rows. The
value of the coarse-graining scale is presented next.
The correlation lengths of the gauge field and the 
radial scalar mode can be inferred from 
the masses at the two minima, given in the next four rows. 
The gauge field stays massless at the origin. We have found
no evidence that it develops a thermal mass. 
The last four rows give the values of the surface tension and
the latent heat of the phase transition, defined according to
\be
\sigma = \int_0^{\phi_{0R}} 
d \phi \sqrt{2 U \left( \rho,T_{cr} \right)}
\label{eighttwo} \ee
and 
\be
\Delta Q = T \left. \frac{\partial}{\partial T}
U \left( \rhz, \tcr \right) \right|_{T=\tcr}.
\label{eightthree} \ee

The well-defined and intuitive picture that we presented in the previous
paragraphs is not valid for all first-order phase transitions. 
In fig. 12 we present the effective average potential for the 
Abelian Higgs model with  
$\lx_R=0.1$, $e^2_R=0.09$ and 
$T^2/\rhz \simeq 2.69$.
We observe several differences with respect to fig. 10.
The discontinuity in the scalar field expectation value
is about three times smaller for the potential of fig. 12. 
The phase transition is more weakly first-order than 
that of fig. 10. 
The other important difference 
is that the potential never 
becomes relatively stable around a certain form. 
During the later stages of the evolution, its outer 
part (for scalar field values
larger than the location of the minimum) starts approaching a 
stable profile, due to the decoupling of the massive modes in this region.
However, over the same time the non-convex part 
starts becoming flatter, as configurations interpolating between the
two minima are being integrated out \cite{convex}. 
The negative curvature at the top of the barrier is expected to
behave $\sim -k^2$ during this stage \cite{convex}. This 
has been verified explicitly through
the analytical integration of the evolution equation for the
$O(N)$-symmetric scalar theory in the large-$N$ limit \cite{largen,analy}. 
The technical reason for this behaviour 
can be traced to the presence of a pole at $w=-1$
for the threshold functions $L^d_n(w)$ \cite{indices,convex}. 
The contributions to the 
evolution equation coming from the scalar fluctuations 
involve threshold functions,
with arguments which include the terms 
$u'_k = U'_k /k^2$ and $u''_k \rht= U''_k \rho /k^2$. 
As the pole cannot be crossed, $U'_k$ and $U''_k$ must go to zero
with $k^2$ in the regions where they are negative. This induces the
flattening of the non-convex parts. 
It is difficult to follow this part of the evolution 
numerically. For this reason we do not display
the full approach to convexity in fig. 12. 
A more thorough presentation, with more detailed figures, 
will be given elsewhere \cite{prep1}.
It is clear, however, that the well-defined separation between the
integration of high-frequency modes and the study of tunnelling
or thermal fluctuations above the barrier
does not exist for the model of fig. 12. 
As a result, one cannot obtain a reliable prediction for the 
nucleation rate by studying only a dominant 
semiclassical configuration. A small variation of the
coarse-graining scale induces large changes in the characteristics
of this configuration and the predicted nucleation rate. 
In fig. 13 we compare various approximations to the potential
at the critical temperature, 
in complete analogy with fig. 11.
The prediction from one-loop perturbation theory 
is that the inclusion of scalar fluctuations 
(line (b)) increases the strength of the first-order phase
transition. This is counter-intuitive and is contradicted by  
the integration of the evolution equation for the potential (line (c)).
The inclusion of the running of the gauge coupling further reduces the 
strength of the phase transition.
Notice that, in figs. 12 and 13, 
we have stopped the evolution at a scale $k$
where the two minima have equal depth for a certain choice of the 
temperature. 
There is considerable arbitrariness in the value of the temperature 
that can be defined as the critical one, and, therefore, 
our choice is not unique. For this reason we have not produced a
table with the characteristics of the 
phase transition. Our present understanding of the dynamics
of weakly first-order phase transitions does not permit
quantitatively precise predictions for their characteristics. 
For recent related work, see ref. \cite{gleiser}.

\setcounter{equation}{0}
\renewcommand{\theequation}{{\bf 9.}\arabic{equation}}

\section*{9. The evolution equations for the $SU(2)$ Higgs model}

We now turn to the discussion of the $SU(2)$ Higgs model, which
displays all the behaviour characteristic of the
electroweak phase transition. 
The inclusion of the $U(1)$ gauge group and fermions introduces only
quantitative modifications, without altering the qualitative 
properties of the phase transition.
The evolution equation for the effective average potential can
be obtained in complete analogy with the intuitive derivation of
section 2. In three dimensions and in terms of the rescaled variables of
eqs. (\ref{threeone}),
it reads
\beq
\frac{\partial u_k}{\partial t} = & 
~-3 u_k + \rht u'_k
\nonumber \\
&~-v_3 L^3_0 \left( u'_k +2 u''_k \rht \right)
-(4N_d-1) v_3 L^3_0 \left( u'_k \right)
- 6 v_3 L^3_0 \left( 2 \et^2 \rht \right).
\label{nineone} \eeq
The only differences in comparison with eq. (\ref{fivefive})
concern the group factors multiplying the contributions of the 
various modes. We consider a model with $N_d$ complex 
doublets, and there are three gauge-group generators.  
Notice also that we employ a rather 
unconventional normalization of the gauge coupling and the gauge field mass.
The rigorous formulation that leads to the above equation,
for a truncated form of the effective average action analogous
to eq. (\ref{twoeight}), can be found in refs. \cite{gauge,abelian,runngauge}.
We find it more convenient to solve numerically the 
evolution equation for the derivative of the potential $u'_k(\rht)$.
This is completely analogous to eq. (\ref{threethree}) with $d=3$ and
the appropriate group factors multiplying the various contributions. 
We neglect the logarithmic running of the gauge coupling in
four dimensions. The $\beta$-function for the three-dimensional running  
of the gauge coupling for the pure $SU(2)$ theory,
in a truncation that assumes the standard form for the gauge field
kinetic term, 
was derived in ref. \cite{gauge}. The generalization for the
$SU(2)$ Higgs model was conjectured in 
ref. \cite{runngauge}. It 
involves a threshold function, 
which suppresses the evolution when the scale $k$ becomes 
smaller than the running mass of the gauge field fluctuations
that drive the evolution. This mass is proportional to the Higgs field
expectation value. 
In section 3, we neglected all threshold behaviour
in eq. (\ref{threefive}) for the running of the gauge coupling. 
However, 
this behaviour will be important in the context of the electroweak phase
transition. The reason is that the $SU(2)$ theory is asymptotically
free and the gauge coupling grows in the infrared. Above a critical
value for the coupling, a confining regime is expected to arise. 
The confining 
regime appears for small Higgs field expectation values, for which 
the gauge field mass is small and the
running of the coupling is not cut off by the threshold function. 
We use the following expression for the running of the rescaled 
gauge coupling 
\be
\frac{d \et^2}{d t} = 
-\et^2 - \frac{4}{3} v_3~88~\et^4 l^3_{NA}
\theta \left( 2 \et^2 \rht \right),
\label{ninetwo} \ee
with
\be
l^3_{NA} = 0.677,
\label{ninethree} \ee
and $v_d$ given by eq. (\ref{twofive}).
This is the pure $SU(2)$ result with an additional 
threshold function, which has been approximated by a theta-function. 
The evolution of $e^2$ is stopped as soon as the scale $k$
becomes smaller than the running mass of the gauge field. 
This results in a $\rho$-dependent running gauge coupling
$e^2$.
There are no fixed points in the evolution of the gauge coupling,
according to eq. (\ref{ninetwo}).
As a result, second-order phase transitions are not expected for 
the $SU(2)$ Higgs model in the region where this equation is 
applicable. In eq. (\ref{ninetwo}) we have neglected the contribution of 
scalar field fluctuations. Their effect is small
and we have preferred to neglect it completely rather than guess
the analogue of the constant $l^3_{NA}$ in their contribution. 
The induced error can be estimated through a comparison with 
the leading-order result from 
the $\epsilon$-expansion 
\be
\frac{d \et^2}{d t} = 
-\epsilon \et^2 - \frac{4}{3} v_4~(88-2 N_d)~\et^4. 
\label{ninefour} \ee
It is a very small effect, $~\sim 2 \%$.
Notice that, if the contribution $\sim N_d$ in eq. (\ref{ninefour})
and the threshold function in eq. (\ref{ninetwo}) 
are neglected, the two equations are in qualitative 
agreement. As eq. (\ref{ninefour}) is derived for $\ex \rightarrow 0$,
the constant $v_d$ takes its value at $d=4$. In contrast, 
eq. (\ref{ninetwo}) has been derived directly in three dimensions.
As a result, $v_3$ appears
in it, along with the constant $l^3_{NA}$.
(For the four-dimensional running one finds $l^4_{NA}=1$ \cite{gauge}.)

The non-zero-temperature effects can be taken into account, in complete
analogy to the discussion in section 4. 
The problem can be reduced to that of integrating the 
three-dimensional evolution equations (\ref{nineone}), (\ref{ninetwo})
for the effective couplings
defined in eqs. (\ref{fivethree}), (\ref{fivefour}). (Notice that
$e^2$ now depends on $k$, $\rho$ and $T$.) 
The evolution starts at the scale $k=T/\thtwo$ (we use $\thtwo=1$)
with the initial condition of eq. (\ref{sixeleven}) for the gauge coupling.
The initial condition for the effective average potential is given by
\beq
U'_{T/\thtwo} (\rho,T)
=~&\lx_R \left\lbrace 
\rho - \rho_{0R} - 
\left[ (4 N_d + 2) + 12 \frac{e^2_R}{\lx_R} \right]
2 v_4 T^2 \left( \frac{2 \sqrt{\pi}}{\thtwo} - \frac{2 \pi^2}{3} \right) 
\right\rbrace 
\nonumber \\
&+U'_{A^0} (\rho,T). 
\label{ninefive} \eeq
The contribution $U'_{A^0} (\rho,T)$ results from the integration of
the fluctuations of the $A^0$ component of the gauge field. 
This component develops a thermal mass and soon decouples from the
evolution, as we have already explained in sections 5 and 6.
For this reason, there are no infrared problems or
non-perturbative dynamics associated with it.
It is, therefore, a good approximation to use the one-loop 
perturbative result for
the contribution of the $A^0$ component 
\beq
U'_{A^0} (\rho,T) =~&
3 \frac{d}{d \rho} \left\lbrace 
\frac{T^4}{2 \pi^2} \int_0^{\infty} dx~x^2 \ln \left( 
1- \exp \left[ - \sqrt{ x^2 + \frac{m^2_L}{T^2} } \right]
\right) \right\rbrace
\nonumber \\
=~& 3 \frac{d}{d \rho} \left( \frac{1}{24} m^2_L T^2 
-\frac{1}{12 \pi} m^3_L T + ... \right),
\label{ninesix} \eeq
with \cite{shapred}
\be
m^2_L = 2 e^2_R \rho + \frac{2(N_d+4)}{3} e^2_R T^2
\label{nineseven} \ee
the thermally corrected mass.

Similarly to the study of the Abelian Higgs model, we can 
check the validity of our approximations by comparing with 
the perturbative predictions.  
We can integrate the evolution equation for the potential, in analogy
to the end of section 6, 
neglecting the running of couplings and threshold effects.
This predicts a critical temperature
in agreement with the lowest-order perturbative result
\cite{sher}
\be
T^2_{cr} = 
\frac{24 \rho_{0R}}{(4 N_d + 2) + 18 \frac{e^2_R}{\lx_R}}.
\label{nineeight} \ee
Corrections to the above expression arise through the 
inclusion of the running of couplings and threshold effects in the
evolution. 
Another check can be made through the comparison with the full 
one-loop perturbative contribution coming from the gauge field, if the
scalar contributions and the running of the gauge coupling are neglected. 
As in section 6, we find a small  
discrepancy, which is due to an imprecise
value for the location of the minimum
at the beginning of the three-dimensional evolution.
This results from the omission of 
threshold effects in the four-dimensional evolution,
and means that our value for the critical temperature 
cannot be trusted to better than 0.3\%. 
In order to eliminate this particular source of uncertainty, 
we determine an
empirical shift for the initial value of the 
minimum  $\delta\rho_0(T/\theta_2)$, 
through a comparison with the perturbative result. 
We incorporate this value
in the initial condition of eq. (\ref{ninefive}), so that 
the perturbative prediction is reproduced, when the
scalar contributions and the running of the gauge coupling are neglected.

\setcounter{equation}{0}
\renewcommand{\theequation}{{\bf 10.}\arabic{equation}}

\section*{10. The electroweak phase transition}

In the previous section we summarized the formalism that we
need for the study of the phase transitions in the $SU(2)$ Higgs
model. We reduced the problem of calculating the 
temperature-dependent effective potential to that of 
integrating the evolution equations (\ref{nineone})
and (\ref{ninetwo}),
for the effective three-dimensional parameters (effective average 
potential and running gauge coupling)
defined in eqs. (\ref{fivethree}), (\ref{fivefour}). 
The evolution starts at the scale $k=T/\thtwo$ (we use $\thtwo=1$),
with the initial conditions given by 
eqs. (\ref{ninefive})--(\ref{nineseven}) and (\ref{sixeleven})
in terms of the renormalized
parameters of the zero-temperature theory.
The temperature-dependent potential is obtained in the limit
$k \rightarrow 0$. 
The initial condition for the potential incorporates the contributions
from the part of the evolution that integrates out fluctuations with
four-dimensional character. The remaining part is purely three-dimensional.
Two algorithms for the numerical integration 
have been presented in detail in ref. \cite{num}.
They give results which 
agree at the 0.3\% level. We expect 
the numerical solution to be an approximation of the solution
of the partial differential equation (\ref{nineone}) with the same 
level of accuracy. 

In fig. 14 we present the evolution 
of the effective average potential $U_k(\rho,T)$ for the 
$SU(2)$ Higgs model with $N_d=1$, as the 
coarse-graining scale $k$ is lowered.
We use the zero-temperature couplings 
$\lx_R=0.02024$, $e^2_R=0.1073$
and choose the temperature $T^2/\rhz \simeq 0.279$, so as to 
be close to the phase transition. 
As we mentioned in the previous section, 
we neglect the logarithmic running of the
four-dimensional couplings that appear in the 
initial conditions for the three-dimensional running. 
It is consistent, therefore, to use the tree-level 
relations for the masses of the Higgs and gauge field, which are 
$m_H=35$ GeV and $m_W=80.6$ GeV.
The corrections to the above values are expected to be small
for the $SU(2)$ Higgs model. However, for the full electroweak
theory they can be sizeable, due to the large Yukawa coupling
of the top quark. 
In fig. 14 we observe an evolution analogous to
that of fig. 10. Initially the potential has only one minimum 
away from the origin, which receives 
a renormalization proportional to 
the running scale $k$ during the evolution. 
At some point, a new, very shallow minimum appears at the origin. 
It is induced by 
the integration of thermal fluctuations, through the
generalization of the Coleman-Weinberg mechanism.
As a result the phase transitions for this parameter range 
are expected to be of first order. 
The evolution slows down at the later stages and the 
potential converges towards a non-convex 
profile. The simultaneous running of the gauge coupling 
$e^2(k,\rho,T)$ is depicted in fig. 15. 
We observe that the gauge coupling increases, because
the $SU(2)$ Higgs model is asymptotically free.
There are no fixed points in the evolution 
for any value of $e^2_R$,  which indicates the 
absence of second-order phase transitions
\footnote{
A second-order phase transition is expected at the point
where the first-order phase transitions are replaced by continuous
crossovers. This happens 
for a Higgs field 
mass larger than the gauge field mass. The description in terms 
of fundamental fields and eq. (\ref{ninetwo})  
cease to be valid at this point. A parametrization of the 
effective average action in terms of composite operators 
is necessary \cite{bound}.}.  
The running of the gauge coupling stops earlier for large $\rho$.
The reason is the decoupling of the massive gauge
field fluctuations when the running scale $k$ becomes 
smaller than their mass $\sqrt{2 e^2 \rho}$. At this point, 
the evolution is automatically stopped by the
theta-function in the r.h.s. of eq. (\ref{ninetwo}).
For values of $\rho$ near the origin 
the magnitude of the gauge coupling continues to increase.
Eventually it reaches a critical value, 
for which a confining regime is expected to emerge.
An estimate for this value is $\alpha=4\et^2/4 \pi=1$.
(The reason for the factor of 4 is our unconventional
normalization of the gauge coupling.)
In figs. 14 and 15, we have stopped the evolution at this
scale, which is determined through the expression
\be
\et^2(k_{conf},\rht=0) = \frac{e^2(k_{conf},\rho=0,T)~T}{k_{conf}} = \pi.
\label{tenone} \ee
The value of $k_{conf}$ has a very weak dependence on the 
value of $\et^2$ for which 
confinement is assumed to set in. The reason is that 
the running coupling diverges very fast at this point of the
evolution. It becomes infinite at the scale 
\be
k_{div}=\frac{e^2_R}{e^2_R \thtwo + 
\left( \frac{4}{3} v_3~88~l^3_{NA} \right)^{-1}} T \simeq 0.1 T,
\label{tentwo} \ee
for $e^2_R=0.1073$ \cite{runngauge,bastian}.
The difference between the scales $k_{conf}$ and $k_{div}$ 
is $\sim 10\%$, independently of the precise definition of $k_{conf}$.

The evolution for $k< k_{conf}$ cannot be described reliably in terms 
of a parametrization such as that of eq. (\ref{twoeight}) for the 
effective average action. Instead, one has to use a basis of
composite operators, corresponding to the bound states that 
are expected to form at these scales \cite{bound}. The two formulations
must be matched at $k= k_{conf}$. 
The various bound states are expected to be massive \cite{symm}, with 
masses of the order of $k_{conf}$, and they should soon decouple. 
The most significant contribution from this last part of the evolution
comes from the appearance of condensates associated with the operator
$F^2$. In ref. \cite{condensate} it was shown that $F^2$ develops a
non-zero expectation value in the confining regime, 
with a reduction of the energy density compared with the 
state with $F^2=0$. This negative contribution affects the potential
of fig. 14. The reason is that the strongly-coupled regime 
appears only near the origin of the potential, where the running
of the gauge coupling has not been cut off.
We expect, therefore, 
a negative contribution to the potential for the region around
$\rho=0$, where the coupling satisfies eq. (\ref{tenone}).
As we have pointed out, the detailed treatment of this part of the evolution
requires extended truncations and a formulation based on composite 
operators. We do not embark here on this 
extensive study, which is the subject of future
work. Instead we use a cruder approximation
for the contribution associated with the $F^2$ condensate.
On dimensional grounds, 
we approximate the negative contribution to the potential by 
\be
\Delta U = - \left( c k_{conf} \right)^3 T.
\label{tenthree} \ee
The constant $c$ is expected to be of order 1. We consider
the values $c=1$ and 0.5 in order to study the effect of this
parameter on the characteristics of the  phase transition. 
In fig. 14 we display the form of the potential, if the 
above contribution with $c=1$ is added to the result of the integration
at $k=k_{conf}$.
This term is added to the potential only 
in the small region around the origin, where the running 
coupling of fig. 15
reaches the value determined by eq. (\ref{tenone}).
As a result, the shallow minimum at the origin becomes as deep as the
minimum at non-zero $\rho$. 
The resulting potential incorporates the dominant effects from
the integration of the fluctuations in the strongly coupled
regime for $k<k_{conf}$.
We have not attempted to account for the $\rho$ dependence of the 
$F^2$ condensate beyond the crude step-like 
behaviour. In ref. \cite{bastian}
this behaviour was smoothed out through the introduction of 
additional phenomenological parameters. As our analysis
cannot provide any hint on the value of these parameters,
we have preferred to neglect them completely. 
This explains the steep rise of the potential of fig. 14 near the origin.
Notice, however, that the influence of this crude approximation 
on the characteristics of the phase transition is rather small.
For example, according to eq. (\ref{eightthree}),
the latent heat is determined by the temperature 
dependence of the energy density at the minimum 
away from the origin. This depends on the magnitude of the
contribution of eq. (\ref{tenthree}), but not on 
its precise $\rho$ dependence. Similarly, the  
surface tension given by eq. (\ref{eighttwo}) involves an 
integration over the whole range between the two minima, which reduces
the effect of our approximation. 
We conclude that the most significant uncertainty in our treatment of
the confining regime is related to the value of the parameter $c$ 
in eq. (\ref{tenthree}).

The potential at the end of the evolution in fig. 14
has the properties of the non-derivative part of a  
coarse-grained free enery, as we discussed in detail 
in section 8. The convergence towards a non-convex profile, 
before the strongly-coupled regime is reached,   
indicates that the various massive modes have already started to decouple. 
The part of the integration until the decoupling of the bound states
is not expected to change this
behaviour, as it is rather short (the mass scale for these states
is set by $k_{conf}$). As a result, we 
expect the separation in two stages that we observed in section 8.
First, the integration of high-frequency 
modes generates a non-convex potential with the 
properties of the non-derivative part of a coarse-grained free energy. 
In this potential, 
tunnelling or thermal fluctuations over the barrier can be studied
through semiclassical techniques. 
In fig. 16 we compare various approximations for the 
form of the potential at
the respective critical temperatures of the first-order phase transitions for
the model of fig. 14. 
Line (a) is the perturbative one-loop result.
For the scalar fluctuations we have used 
expressions analogous to eq. (\ref{ninesix}), with thermally corrected masses
given by 
\beq
m^2_G = &\lx_R \rho + 
\left( \frac{3}{4}e^2_R + \frac{2 N_d +1}{12}\lx_R \right)
\left(T^2 - \frac{12 \rho_{0R}}{(2 N_d + 1) + 9 \frac{e^2_R}{\lx_R}}
\right) 
\nonumber \\
m^2_R = &3 \lx_R \rho + 
\left( \frac{3}{4}e^2_R + \frac{2 N_d +1}{12}\lx_R \right)
\left(T^2 - \frac{12 \rho_{0R}}{(2 N_d + 1) + 9 \frac{e^2_R}{\lx_R}}
\right) 
\label{tenfour} \eeq
for the Goldstone and radial modes respectively.
Line (b) results from the integration of the evolution equation 
if the running of $e^2$ is neglected.
We observe that the strength of the first-order phase
transition is reduced. We have observed the same behaviour in section 8.
The proper integration of scalar fluctuations through the renormalization
group reduces the size of the barrier, contrary to 
the predictions of perturbation theory. 
Line (c) results from the integration of the evolution equation 
for the potential
with the running of $e^2$ included, but neglecting the effect of the
$F^2$ condensate.
The presented potential corresponds to the coarse-graining 
scale $k_b$, where the
evolution slows down, in complete analogy to the discussion in section 8.
As the first-order phase
transition is triggered by the gauge field fluctuations, its
strength is increased compared with line (b), 
because the ``effective'' gauge coupling 
is larger than $e^2_R$. 
Line (d) incorporates the contribution of eq. (\ref{tenthree}) with
$c=1$, which 
comes from the $F^2$ condensate. We observe a dramatic increase 
of  the strength of the first-order phase transition. 
Line (e) is similar to line (d), but for a value $c=0.5$. The
influence of the condensate diminishes, but again an increase
of the strength of the phase transition is observed.
We should also mention that this increase 
is observed in higher orders of perturbation theory, where the
running of the couplings is partly accounted for \cite{twoloop}.

The characteristics of the phase transition can be deduced from
the form of the potential, in complete analogy to the discussion of
section 8. 
In table 3 we summarize them
for the model of fig. 14. The three columns correspond to 
the curves (c), (d) and (e) of fig. 16.
The critical temperature and the discontinuity 
in the order parameter are given in the first three rows. The
value of the scale $k_{conf}$
where confinement sets in is presented next.
We do not give a value for $k_{conf}$ in the first column, because
the effects of the strongly-coupled regime are not taken into account
for the potential of line (c). 
The correlation lengths of the gauge field and the 
radial scalar mode at the minimum away from the origin
can be inferred from 
the masses, given in the next three rows. 
We do not give any values for the masses at the origin. 
There, the proper description should include bound states instead
of fundamental fields, and is the subject of future work.
We have given the value of the mass of the radial scalar mode for
two scales. The reason is that 
the running of the quartic coupling 
(which determines this mass)
never stops. This is 
due to the presence of Goldstone modes at the minimum away from the
origin, which
never decouple.
In contrast, the running of the gauge coupling (which determines the mass
of the gauge field) is stopped by the threshold function in 
eq. (\ref{ninetwo}).
For line (e), the confinement scale where we stop the
evolution is larger than the mass of the radial mode.
The last four rows give the values of the surface tension and
the latent heat of the phase transition, defined according to
eqs. (\ref{eighttwo}) and (\ref{eightthree}),
respectively.
The influence of the confining regime 
on the characteristics of the phase transition is apparent.
Also, the dependence on 
the parameter $c$ in eq. (\ref{tenthree})
is significant. We can derive a phenomenologically 
motivated value for this constant by comparing with
the results of other approaches for the quantities of 
table 3.

The values of table 3 for the critical temperature and the discontinuity 
in the Higgs field expectation value are in good agreement with the
results of ref. \cite{bastian}, obtained for the same
approximations for the quantity of eq. (\ref{tenthree}). 
A more crucial test is provided by comparison with the
results from lattice studies.
Unfortunately, the studies of refs. \cite{kaj,fodor}
where carried out for different electroweak parameters. 
In ref. \cite{kaj} different masses for the Higgs field
were used. In ref. \cite{fodor} 
the value of the renormalized gauge coupling is 
different. This results in a different value for the minimum
of the potential for the zero-temperature theory than the one we used.
The direct comparison is in progress \cite{prep2}. 
However, there is an indirect way for an immediate comparison. 
In ref. \cite{compar} the lattice results were compared with the
predictions of two-loop perturbation theory. According to figs. 7 and 8
of this reference, good agreement 
is observed for Higgs field masses in the 20--50 GeV range. 
This agreement cannot be explained by our analysis. Our fig.
16 indicates that the non-perturbative effects of the
strongly-coupled regime 
are substantial already at $m_H=35$ GeV.
In spite of that, we can use the perturbative 
results of ref. \cite{compar}
as a phenomenological fit to the lattice data. 
We deduce the values 
$\tcr/m_H \simeq 2.76$ 
and 
$\Delta Q/\tcr^{4} \simeq 0.103$ 
from figs. 4 and 6 in ref. \cite{compar} 
\footnote{
The quantity 
$\phi_0(\tcr)/\tcr \simeq 1.26$, which can be deduced form fig. 5, 
corresponds to the 
expectation value of the operator $\phi^* \phi$. 
The values quoted in our table 3 correspond to expectation values
of $\phi$ in a formalism with gauge fixing.}.
They can be compared with the results presented in table 3. 
We observe that they are in good agreement with column (d). 
This agreement is also observed through the direct comparison with the
lattice results, for Higgs field masses in the 30--50 GeV range \cite{prep2}.
We conclude that the characteristics of the first-order
phase transition are properly reproduced by our analysis,
if the non-perturbative contribution from the 
strongly-coupled regime in the evolution of the potential
is approximated by eq. (\ref{tenthree}) with $c=1$. 
The non-perturbative effects dramatically increase the 
strength of the phase transition. Their effect is much larger
than the effect due to the growth of the running gauge coupling.
We should point out, however, that the determination of the value
$c=1$ through an explicit calculation in the context of the
effective average action is still pending.

We now turn to the question of the nature of the phase
transition for large Higgs field masses. In fig. 17
we plot the effective average potential $U_k(\rho,T)$ 
as the 
coarse-graining scale $k$ is lowered. 
The Higgs and gauge field masses are 
$m_H=70$ GeV and $m_W=80.6$ GeV respectively, and the temperature 
$T^2/\rhz \simeq 0.882$.
We observe the running of the minimum and 
the curvature at the origin becoming less negative. 
In the region around the origin, the running gauge coupling
reaches the critical value of eq. (\ref{tenone}), 
at which the confining regime is 
expected to set in. The size of this region relative to the 
location of the minimum of the potential is larger than in the case of 
fig. 14. 
Confinement sets in while the curvature at the origin
is still negative. At this point we stop the evolution and 
add the contribution of eq. (\ref{tenthree}) 
to the potential, in the region around the origin where the coupling
has the critical value. (We display the evolution with $c=0.5$, but
we give the characteristics of the transition for both $c=1$ and 0.5.)
This results in a new minimum between the origin 
and the original minimum of the potential. 
A first-order phase transition is predicted,
but both minima 
now exist at non-zero values of the Higgs field. 
This is the major difference between 
figs. 17 and 14. Other differences concern
the strength of the predicted first-order phase transition. The discontinuity
in the Higgs field expectation value is smaller in fig. 17. 
Also the potential does not converge towards a stable profile, which
would indicate the decoupling of massive modes. This is the 
typical behaviour of weakly first-order
phase transitions, which we have already discussed in detail in 
section 8. As we have mentioned at the end of that section,
the dynamics of such phase transitions is not well understood.
In table 4, we list the characteristics which are predicted
by the potential at the end of the evolution in fig. 17, in complete
analogy to table 3. The location of the new minimum 
is also included in the table. 
The latent heat is calculated through
\be
\Delta Q = T \frac{\partial}{\partial T}
\left\lbrace U \left( \rhz, \tcr \right) -
U \left( \rho_{np}, \tcr \right) 
\rule{0mm}{5mm} 
 \right\rbrace_{T=\tcr}.
\label{tenfive} \ee
We point out that,
similarly to section 8, 
there is considerable arbitrariness in defining  
which temperature is 
the critical one, and, therefore, 
the values of table 4 must be viewed as indicative, rather than
quantitatively precise. 
The appearance of a minimum at a non-zero Higgs expectation value has 
also 
been observed in ref. \cite{bastian} and within the gap-equation approach in 
ref. \cite{owe}. 

For even larger Higgs masses a qualitative change 
is expected. The evolution resembles that in fig. 17, until the 
gauge coupling reaches its critical value. 
When the strongly-coupled regime sets in, its extent is larger
than the location of the minimum of the potential. 
This indicates that there is no longer any minimum where the 
description in terms of fundamental fields is possible. 
The transition to a 
non-perturbative vacuum is expected to 
be continuous, without the appearance of singularities. 
As a result there is no phase transition any more. This indicates
the change from first-order phase transitions
to analytical crossovers for large Higgs field masses. Our analysis
predicts
a critical Higgs field mass in the range 80--100 GeV. 
The possibility of a crossover was suggested in ref. \cite{runngauge}. It 
is supported by the studies of ref. \cite{bastian}, where arguments
similar to ours were given,  and ref. \cite{owe}, where the gap-equation 
approach was followed. The most reliable results that confirm this
possibility have been obtained through the lattice approach \cite{crosslat}.

\setcounter{equation}{0}
\renewcommand{\theequation}{{\bf 11.}\arabic{equation}}

\section*{11. Conclusions}

In this paper we discussed the high-temperature phase transitions 
for the Abelian and $SU(2)$  Higgs models by employing the 
method of the effective average action.
Our aim was to present a method that can deal with the most important 
problems to be encountered in the study of the electroweak phase
transition. We identified as such 
the problem of infrared divergences 
in the perturbative approach and the absence of a coarse-graining 
scale in the calculation of potentials, which are used for
the study of first-order phase transitions. 
The method of the effective average action 
provides a resolution
of these problems through the introduction of an infrared cutoff
scale $k$. This 
can be identified with the coarse-graining scale. The effective 
average potential 
has the characteristics of the non-derivative part 
of a coarse-grained free
energy, on which a proper treatment of statistical systems 
can be based. It does not have to be convex and 
it is the most natural tool for the study of first-order phase transitions.
The resolution
of the problem of infrared divergences 
is achieved through the use of the renormalization group, which 
is built in the method. The dependence 
of the effective average action on the scale $k$
is described by an exact renormalization-group equation. 
From this, evolution equations for the potential and other invariants
can be derived. The integration of the evolution equations 
determines all the couplings of 
the renormalized theory at zero and non-zero temperature
without the appearance of any non-physical divergences.
The reduction of the effective dimensionality of the system from four to 
three, at scales $k$ smaller than the temperature, is easily demonstrated.
The complete phase diagram can be determined without the 
need to resort to lattice simulations, which require 
long computer time and do not provide intuition on the nature 
of the physical behaviour. Moreover, the discussion of the 
phase diagram in terms of the evolution of the potential
provides information on the running of all the generalized couplings
of the theory (the 1PI Green's functions at zero external momenta).
 It also gives a detailed picture of first-order
phase transitions, for which the shape of the whole potential, and not
the running of a few couplings, is required. 
The third important problem in the study of the electroweak phase
transition concerns the strongly-coupled
regime in the symmetric phase of the electroweak theory. 
The possibility for a proper treatment is provided within this 
approach, through the use of composite operators for the parametrization
of the effective average action in this regime. 
We did not embark here on this 
extensive study, which is the subject of future
work. Instead we relied on a cruder determination of the
contribution to the potential coming from
the effects of the confining regime. 
Through the integration of the evolution equations  
we determined the potential at the scale $k_{conf}$, where confinement 
is expected to set in and an $F^2$ condensate appears.
The effect on the potential was determined from this scale through
a dimensional argument, which introduces an undetermined parameter.
This was fixed through the comparison of our results with those of the
lattice approach.

We found that the Abelian Higgs model with 
a number of complex scalar fields $N_c \geq 5$
has a region of second-order
phase transitions and a region of first-order ones, which are 
separated by a tricritical point. 
The second-order phase transitions are governed by two
fixed points: the Wilson-Fisher fixed point and the
Abelian one.
We determined the universal form of the potential
at the fixed points and the tricritical one, 
from which the values of all the 
generalized couplings can be obtained (by taking derivatives
with respect to the field). Moreover, we investigated through
the evolution of the potential the relative stability of 
the fixed points. We determined critical exponents and crossover
curves, which parametrize the physical 
behaviour near the critical temperature. 
For $N_c < 5$ the fixed points disappear and only first-order phase
transitions exist. Our determination of 
the critical $N_c$ for the qualitative change of behaviour
is subject to uncertainties, due to the truncated form of 
the effective average action that we 
used. We can firmly establish that 
$\left( N_c \right)_{cr} = {\cal{O}}(1)$. However, 
the existence of 
the region of second-order phase transitions 
for $N_c=1$ is not excluded. 
This leaves open the possibility of a second-order phase transition
for low-temperature superconductors, which belong to the
same universality class as the Abelian Higgs model. 
The first-order phase transitions were studied in detail.
We presented an explicit realization of a coarse-grained 
potential for a strongly first-order phase transition, 
from which its characteristics were derived. We showed that
the study of such phase transitions can be naturally separated
in two parts: First the high-frequency modes are integrated out
and a non-convex potential is generated (often through the
Coleman-Weinberg mechanism of radiative symmetry breaking).
At a second stage,
the fluctuations that drive the phase transition
(instantons or critical bubbles) can be studied with semiclassical 
techniques, using the potential. 
This separation is impossible for weakly first-order
phase transitions. The two parts merge, and the dominant
semiclassical configurations do not provide a complete 
description. 

The $SU(2)$ Higgs model with one scalar doublet 
(which exhibits the 
same qualitative behaviour as the electroweak theory)
has only first-order phase transitions
for Higgs field masses smaller than the 
gauge field mass. 
The gauge coupling near the origin of the potential 
grows as $k$ is lowered, until it reaches
a critical value, for which a confining regime is expected to
set in. At this point an $F^2$ condensate emerges with a
negative contribution to the potential around the origin. This 
increases dramatically the strength of the first-order phase transition.
We determined its characteristics 
for $m_H=35$ GeV, which are in good agreement with two-loop
perturbation theory and lattice results. 
We also considered the case $m_H=70$ GeV, for which the
first-order phase transition is much weaker.
The behaviour that we described above in the context of
the Abelian Higgs model for such transitions was again observed. 
For Higgs masses above 80--100 GeV we no longer found a 
two-minimum potential with a prediction of a first-order transition.
Instead we found indications that an analytical crossover 
connects the two regions of the phase diagram that correspond
to the symmetric phase and the phase with spontaneous symmetry breaking.

The picture that emerged from our study of the
high-temperature phase transitions
for the Abelian and $SU(2)$ Higgs models has a rich structure and
provides physical intuition as well as quantitative information.
The method of the effective average action has also been applied to 
pure scalar theories (a list of references is given in the introduction).
The second-order phase transitions of the $O(N)$-symmetric theory 
have been
discussed in terms of the evolution of the potential, with the
identification of fixed points. Critical exponents and amplitudes, as 
well as the full critical equation of state, have been computed.
Two-scalar theories have also been considered, which exhibit
a richer fixed-point structure, crossover phenomena and first-order
phase transitions. 
The combined picture of phase transitions 
for high temperature field theories incorporates a wide 
range of physical behaviours. The most important pending
problem in this approach is the use of a truncated ansatz for the
effective average action. An intrinsic check of the accuracy of the
predictions can be achieved through the calculation of the change
that a more general ansatz induces to the results. 
A systematic way of carrying out this check 
has not been firmly established yet. This is the most important
direction for further work, which will lead to 
improved quantitative accuracy.

\underline{Acknowledgements}: We  would like to thank
Z. Fodor, M. Shaposhnikov and C. Wetterich for discussions.

\newpage

\setcounter{equation}{0}
\renewcommand{\theequation}{{\bf F.}\arabic{equation}}

\section*{Figures}

\renewcommand{\labelenumi}{Fig. \arabic{enumi}}
\begin{enumerate}

\item  
The effective potential for the 
Abelian Higgs model with $N_c=1$ in the approximation
that scalar fluctuations are neglected.
$\lx_R=0.02$, $e^2_R=0.09$,
$T^2/\rhz \simeq 0.953$.
Line (a) is the perturbative one-loop result.\\ 
Line (b) is the result of the numerical integration of the 
evolution equation without the scalar contributions.\\
Line (c) is the result of the numerical integration if the initial value for 
the minimum of the
potential is shifted by a relative amount $\simeq 0.3 \%$.
\vspace{5mm}

\item  
The phase diagram for the Abelian Higgs model with $N_c=250$.
The evolution equations (\ref{fivesix}), 
(\ref{threeeight}) with $d=3$ are
used. 
The critical value for the existence of three fixed points
is $\left( N_c \right)_{cr} = 222$. 
\\ {\bf Correction: $\tilde{\lx}$, $\tilde{e}^2$ instead of $\lx$, $e^2$.
The direction of flows is described in the text.}
\vspace{5mm}

\item  
The running $u'(\rht)$ for the 
Abelian Higgs model with $N_c=5$. 
$\lx_R=0.5$, $e^2_R=10^{-6} \times e^2_{A}$,
$T^2/\rhz \simeq 2.10$.
The system approaches first the Wilson-Fisher fixed point and subsequently
the Abelian fixed point. The final running leads to the phase with 
spontaneous symmetry breaking.
\\ {\bf Correction: $\tilde{\rho}$ instead of $\rho$.}
\vspace{5mm}

\item  
The evolution of $\kx$, $\lx$ and $e^2$ for the 
model of fig. 3. The approach to the two fixed points is
apparent. 
\\ {\bf Correction: $\tilde{\lx}$, $\tilde{e}^2$ instead of $\lx$, $e^2$.}
\vspace{5mm}

\item  
The ``effective'' exponent $\nu$ 
as the critical temperature is approached for the model of fig. 3.
\vspace{5mm}

\item  
The Wilson-Fisher (WF), Abelian (A) and tricritical (T) fixed point,
and the inflection point (I), for $N_c=5$.
\\ {\bf Correction: $\tilde{\rho}$ instead of $\rho$.}
\vspace{5mm}

\item  
The phase diagram for the Abelian Higgs model with $N_c=1$.
The evolution equations (\ref{fivesix}), 
(\ref{threeeight}) with $d=3$ are
used.  
\\ {\bf Correction: $\tilde{\lx}$, $\tilde{e}^2$ instead of $\lx$, $e^2$.
The direction of flows is described in the text.}
\vspace{5mm}

\item  
The running $u'(\rht)$ for the 
Abelian Higgs model with $N_c=1$. 
$\lx_R=0.5$, $e^2_R=10^{-7} \times e^2_{A}$,
$T^2/\rhz \simeq 6.41$.
The system first approaches the Wilson-Fisher fixed point. Subsequently
a new minimum appears at the origin, which eventually becomes the
absolute minimum of the potential.
\\ {\bf Correction: $\tilde{\rho}$ instead of $\rho$.}
\vspace{5mm}

\item  
The evolution of $\kx$, $\lx$ and $e^2$ for the 
model of fig. 8. 
\\ {\bf Correction: $\tilde{\lx}$, $\tilde{e}^2$ instead of $\lx$, $e^2$.}
\vspace{5mm}

\item  
The effective average potential $U_k(\rho,T)$ for the 
Abelian Higgs model with $N_c=1$ as the 
coarse-graining scale $k$ is lowered.
$\lx_R=0.02$, $e^2_R=0.09$,
$T^2/\rhz \simeq 0.841$.
\vspace{5mm}

\item  
The potential at the critical temperature in various approximations 
for the model of fig. 10. \\
Line (a) is the perturbative one-loop result if the 
scalar fluctuations are neglected. \\
Line (b) is the perturbative one-loop result with the 
scalar fluctuations included. \\
Line (c) results from the integration of the evolution equation 
if the running of $e^2$ is neglected. \\
Line (d) results from the integration of the evolution equation 
with the running of $e^2$ included.
\vspace{5mm}

\item  
Same as fig. 10 for
$\lx_R=0.1$, $e^2_R=0.09$,
$T^2/\rhz \simeq 2.69$.
\vspace{5mm}

\item 
Same as fig. 11 for the model of 
fig. 12. 
\vspace{5mm}

\item  
The effective average potential $U_k(\rho,T)$ for the 
$SU(2)$ Higgs model with $N_d=1$ as the 
coarse-graining scale $k$ is lowered.
The couplings correspond to $m_H=35$ GeV, $m_W=80.6$ GeV
($\lx_R=0.02024$, $e^2_R=0.1073$).
$T^2/\rhz \simeq 0.279$.
\vspace{5mm}

\item  
The running gauge coupling $e^2(k,\rho,T)$ for the 
model of fig. 14. 
\vspace{5mm}

\item  
The potential at the critical temperature in various approximations 
for the model of fig. 14. \\
Line (a) is the perturbative one-loop result. \\
Line (b) results from the integration of the evolution equation 
if the running of $e^2$ is neglected. \\
Line (c) results from the integration of the evolution equation 
if the running of $e^2$ is included, but the effect of
the $F^2$ condensate is neglected. \\
Lines (d) and (e) result from the integration of the evolution equation 
if both the running of $e^2$ and the effect of
the $F^2$ condensate (in two different approximations)
are included.
\vspace{5mm}

\item  
The effective average potential $U_k(\rho,T)$ for the 
$SU(2)$ Higgs model with $N_d=1$ as the 
coarse-graining scale $k$ is lowered.
$m_H=70$ GeV, $m_W=80.6$ GeV
($\lx_R=0.08097$, $e^2_R=0.1073$).
$T^2/\rhz \simeq 0.882$.

\end{enumerate}

\newpage

\section*{Tables} 

\begin{table} [h]
\renewcommand{\arraystretch}{1.5}
\hspace*{\fill}
\begin{tabular}{|c||c|c|c|c|c|c|}     \hline
$N_c$ & 100 & 10 & 8 & 6 & 5
\\ \hline 
$\nu_A$ & 0.989 & 0.894 & 0.865 & 0.810 & 0.759
\\ \hline
\end{tabular}
\hspace*{\fill}
\renewcommand{\arraystretch}{1}
\caption[y]
{
The critical exponent $\nu_A$ for the Abelian fixed point and various values 
of $N_c$.} 
\end{table}

\begin{table} [h]
\renewcommand{\arraystretch}{1.5}
\hspace*{\fill}
\begin{tabular}{|c|c|}     
\hline
$\tcr/\sqrt{\rhz}$ & 0.917 
\\ \hline 
$\phi_{0R}(\tcr)/\sqrt{\rhz}$ & 0.235
\\ \hline 
$\Delta \phi(\tcr)/\tcr$ & 0.256
\\ \hline 
$k_b/\sqrt{\rhz}$ & $0.985 \times 10^{-2}$
\\ \hline 
$m_W(0,\tcr)/\sqrt{\rhz}$ & 0
\\ \hline 
$m_W \left( \rhz,\tcr \right) /\sqrt{\rhz}$ & $0.698 \times 10^{-1}$
\\ \hline 
$m_H(0,\tcr)/\sqrt{\rhz}$ & $0.143 \times 10^{-1}$
\\ \hline 
$m_H \left( \rho_{0R},\tcr \right) /\sqrt{\rhz}$ 
& $0.146 \times 10^{-1}$
\\ \hline 
$\sigma/\rhz^{3/2}$ & $0.129 \times 10^{-3}$
\\ \hline 
$\sigma/\tcr^{3}$ & $0.167 \times 10^{-3}$
\\ \hline 
$\Delta Q/\rhz^{2}$ & $0.112 \times 10^{-2}$
\\ \hline 
$\Delta Q/\tcr^{4}$ & $0.158 \times 10^{-2}$
\\ \hline 
\end{tabular}
\hspace*{\fill}
\renewcommand{\arraystretch}{1}
\caption[y]
{
Characteristics of the first-order phase transition 
for the Abelian Higgs model with $N_c=1$, $\lx_R=0.02$, $e^2_R=0.09$.} 
\end{table}

\newpage

\begin{table} [h]
\renewcommand{\arraystretch}{1.5}
\hspace*{\fill}
\begin{tabular}{|c||c|c|c|}     
\hline
~ & (c) & (d) & (e)
\\ \hline \hline
$\tcr$/GeV 
& 93.7 & 91.8 & 93.5
\\ \hline 
$\phi_0(\tcr)$/GeV 
& 92.4 & 116 & 96.5
\\ \hline 
$\Delta \phi(\tcr)/\tcr$ 
& 0.986 & 1.27 & 1.03
\\ \hline 
$k_{conf}$/GeV 
& --- & 11.7 & 11.9
\\ \hline 
$m_W \left( \rhz,\tcr \right)$/GeV 
& 33.9 & 41.0 & 35.1
\\ \hline 
$m_H \left( \rho_0,\tcr,k=m_H \right)$/GeV
& 7.83 & 12.4 & ---
\\ \hline 
$m_H \left( \rho_0,\tcr,k=k_{conf} \right)$/GeV
& --- & 12.3 & 8.68
\\ \hline 
$\sigma$/(GeV$)^3$ 
& $1.27 \times 10^{4}$ & $3.04 \times 10^{4}$ & $1.41 \times 10^{4}$
\\ \hline 
$\sigma/\tcr^{3}$ 
& $1.54 \times 10^{-2}$ & $3.93 \times 10^{-2}$ & $1.73 \times 10^{-2}$
\\ \hline 
$\Delta Q$/(GeV$)^4$ 
& $5.46 \times 10^{6}$ & $8.21 \times 10^{6}$ &  $5.81 \times 10^{6}$  
\\ \hline 
$\Delta Q/\tcr^{4}$ 
& $7.08 \times 10^{-2}$ & $1.16 \times 10^{-1}$ & $7.60 \times 10^{-2}$ 
\\ \hline 
\end{tabular}
\hspace*{\fill}
\renewcommand{\arraystretch}{1}
\caption[y]
{
Characteristics of the first-order phase transition 
for the $SU(2)$ Higgs model with $N_d=1$, 
$m_H=35$ GeV, $m_W=80.6$ GeV.} 
\end{table}

\newpage

\begin{table} [h]
\renewcommand{\arraystretch}{1.5}
\hspace*{\fill}
\begin{tabular}{|c||c|c|c|}     
\hline
~ & (c) & (d) & (e)
\\ \hline \hline
$\tcr$/GeV 
& 165 & 156 & 163
\\ \hline 
$\phi_0(\tcr)$/GeV 
& 68.0 & 123 & 88.3
\\ \hline 
$\phi_{np}(\tcr)$/GeV 
& 0 & 32.1 & 38.1  
\\ \hline 
$\Delta \phi(\tcr)/\tcr$ 
& 0.412 & 0.583 & 0.308
\\ \hline 
$k_{conf}$/GeV 
& --- & 19.9 & 20.8  
\\ \hline 
$m_W \left( \rhz,\tcr\right)$/GeV 
& 30.8 & 46.4 & 36.6  
\\ \hline 
$m_H \left( \rho_0,\tcr,k=m_H \right)$/GeV
& 8.70 & 29.8 & ---
\\ \hline 
$m_H \left( \rho_0,\tcr,k=k_{conf} \right)$/GeV
& --- & 29.6 & 18.8  
\\ \hline 
$\sigma$/(GeV$)^3$ 
& $1.03 \times 10^{4}$ & $ 7.89 \times 10^{4}$ & $ 1.54 \times 10^{4}$
\\ \hline 
$\sigma/\tcr^{3}$ 
& $2.29 \times 10^{-3}$ & $ 2.08 \times 10^{-2}$ & $ 3.56 \times 10^{-3}$
\\ \hline 
$\Delta Q$/(GeV$)^4$ 
& $1.13 \times 10^{7}$ & $ 3.02 \times 10^{7}$ &  $ 1.56 \times 10^{7}$  
\\ \hline 
$\Delta Q/\tcr^{4}$ 
& $1.52 \times 10^{-2}$ & $ 5.10 \times 10^{-2}$ & $ 2.21 \times 10^{-2}$ 
\\ \hline 
\end{tabular}
\hspace*{\fill}
\renewcommand{\arraystretch}{1}
\caption[y]
{
Characteristics of the first-order phase transition 
for the $SU(2)$ Higgs model with $N_d=1$, 
$m_H=70$ GeV, $m_W=80.6$ GeV.} 
\end{table}

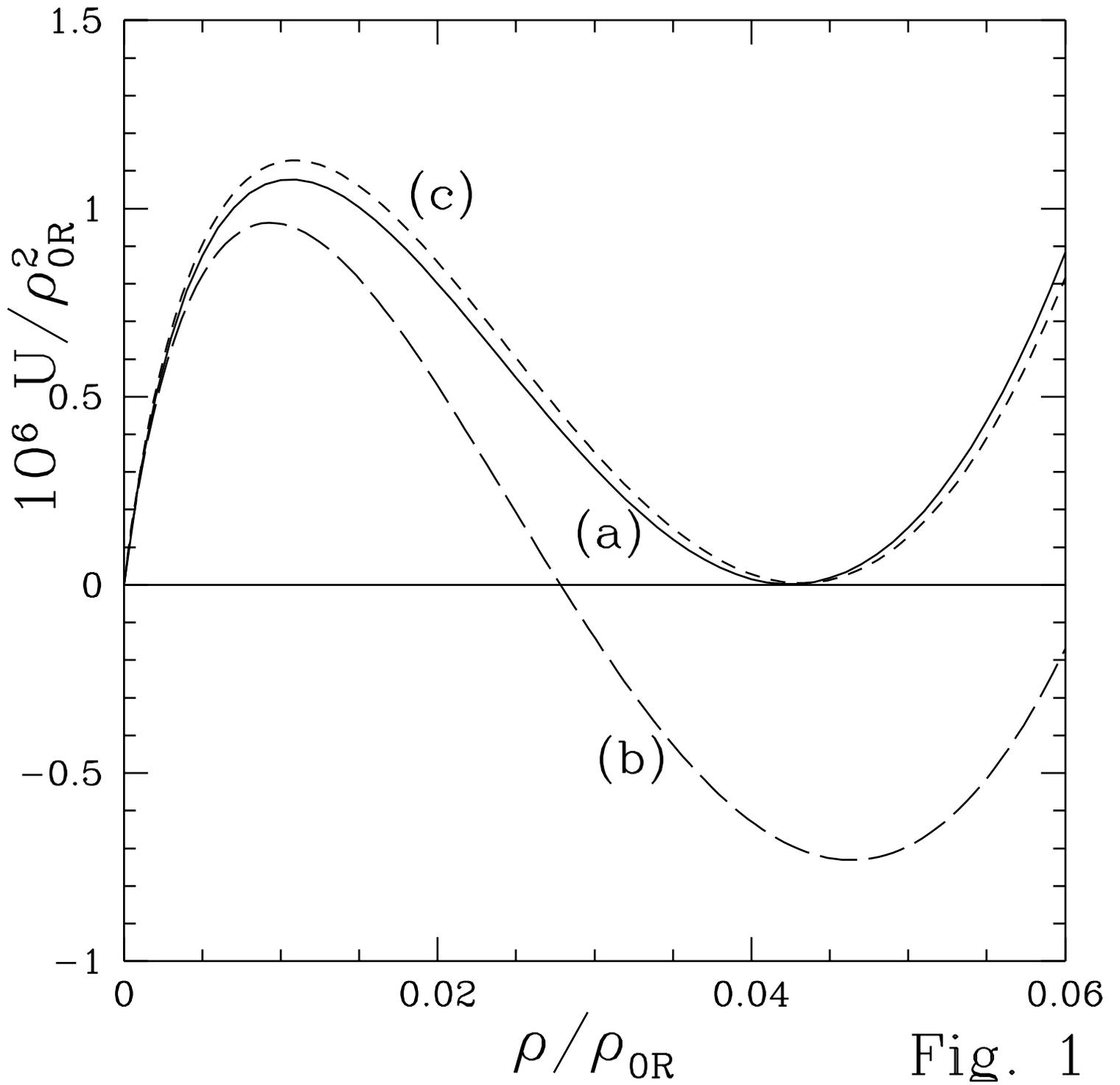
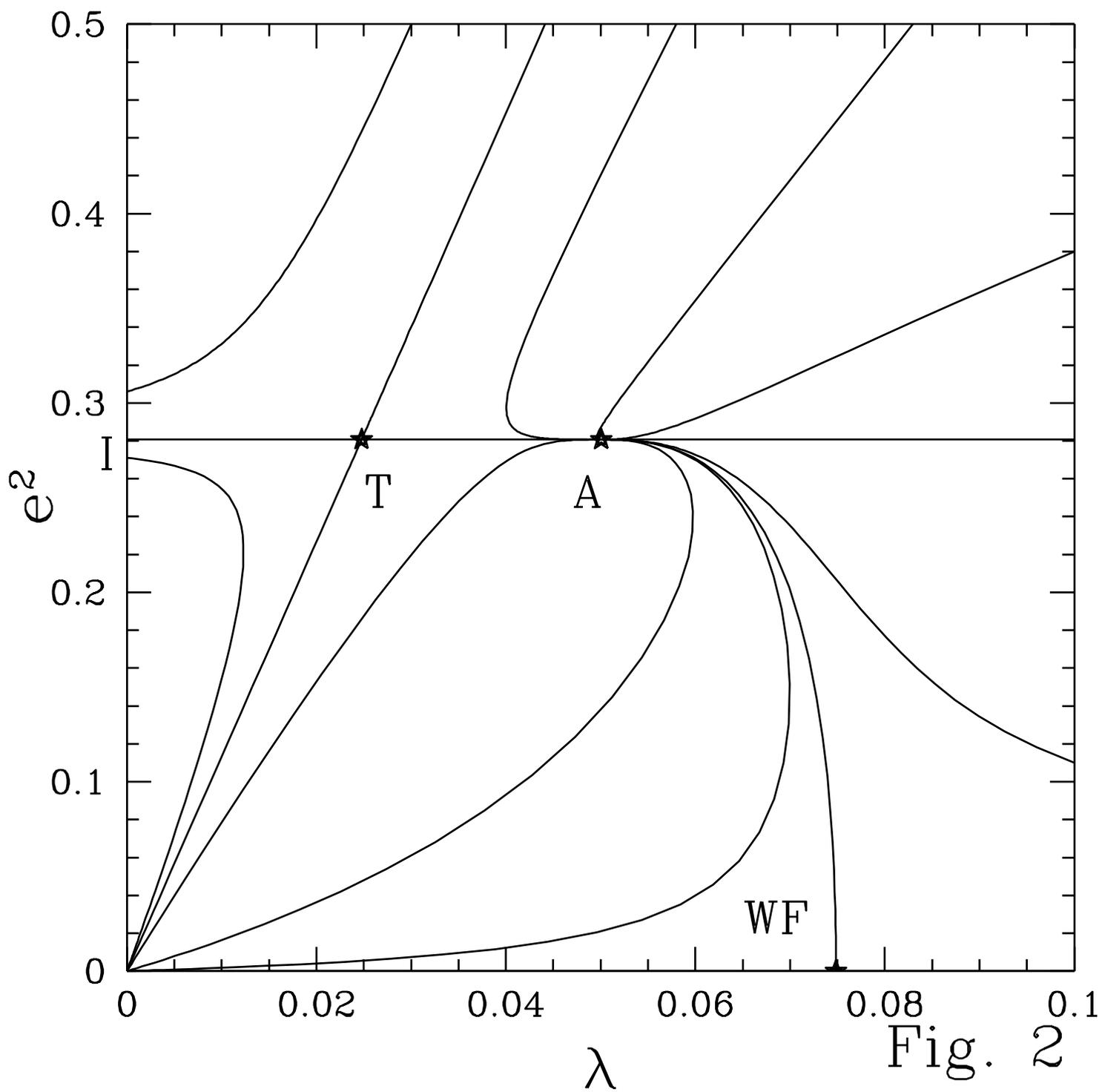
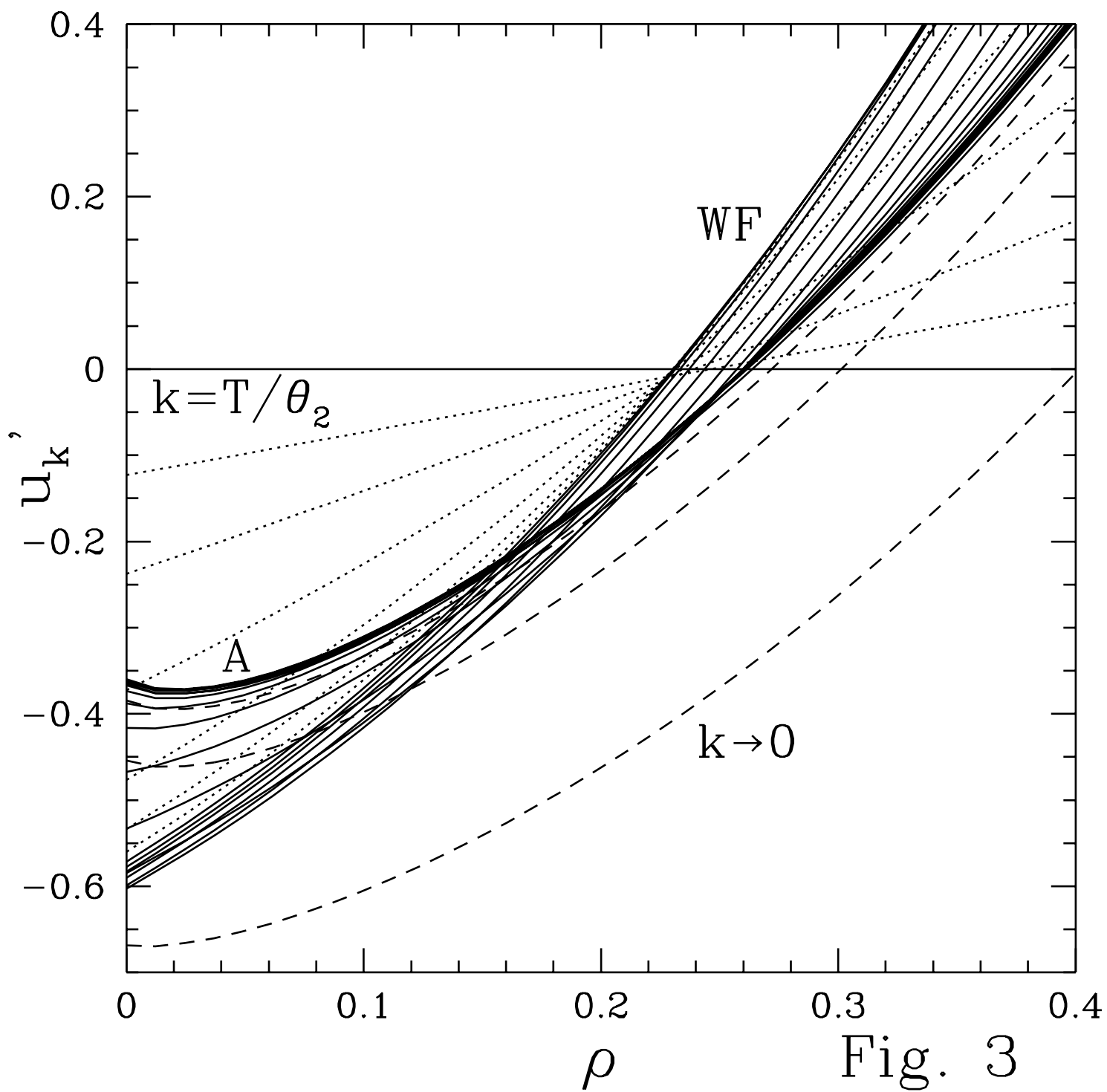
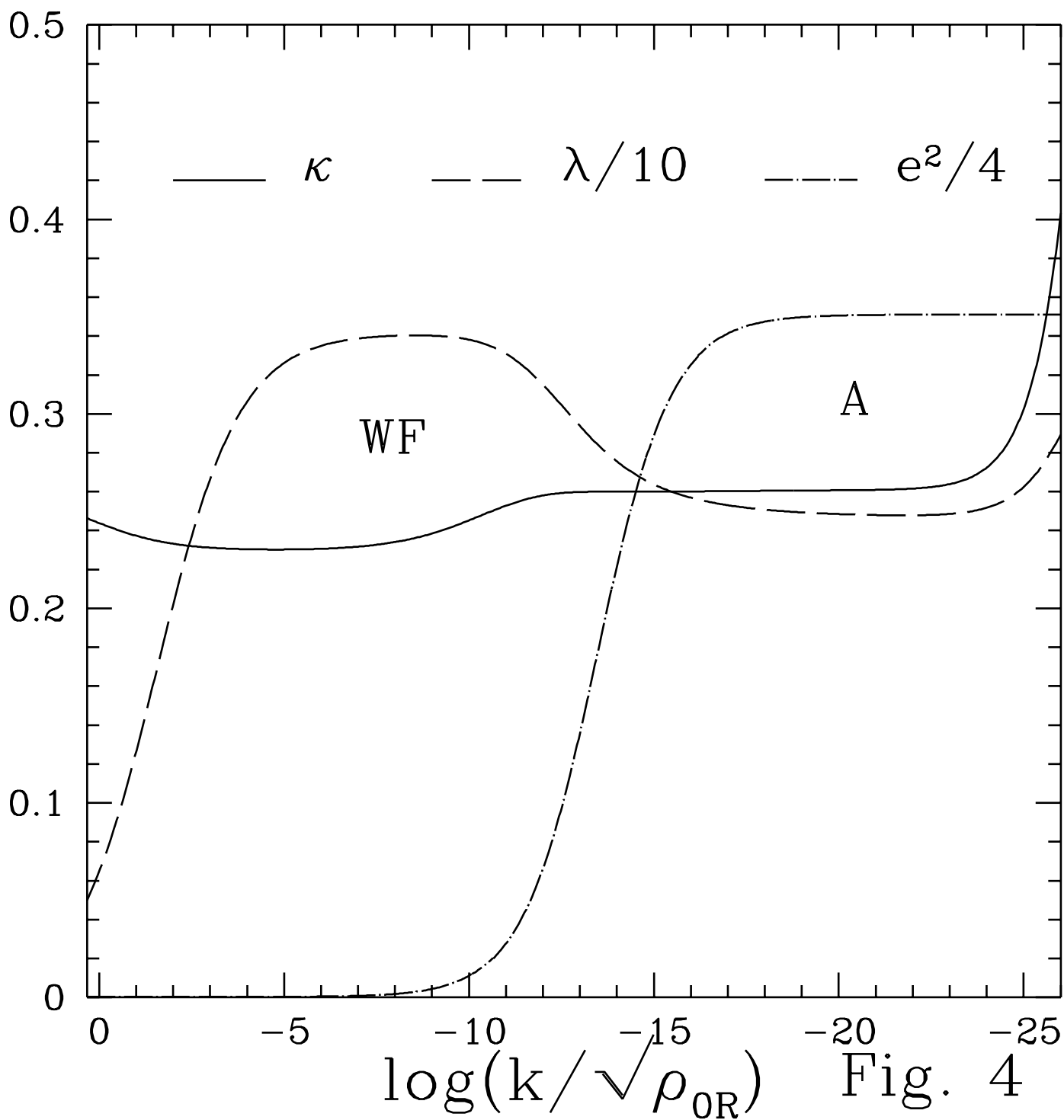
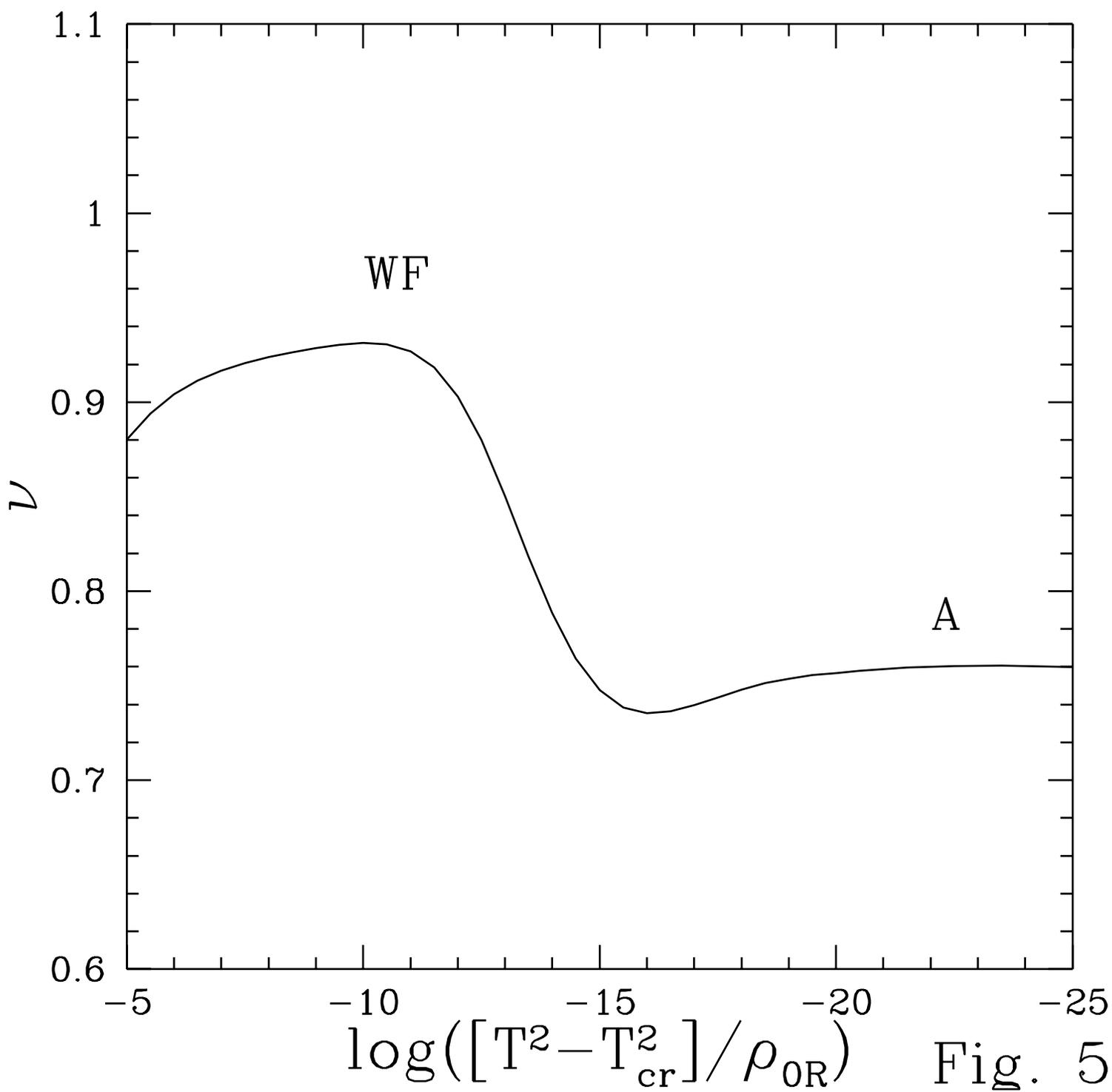
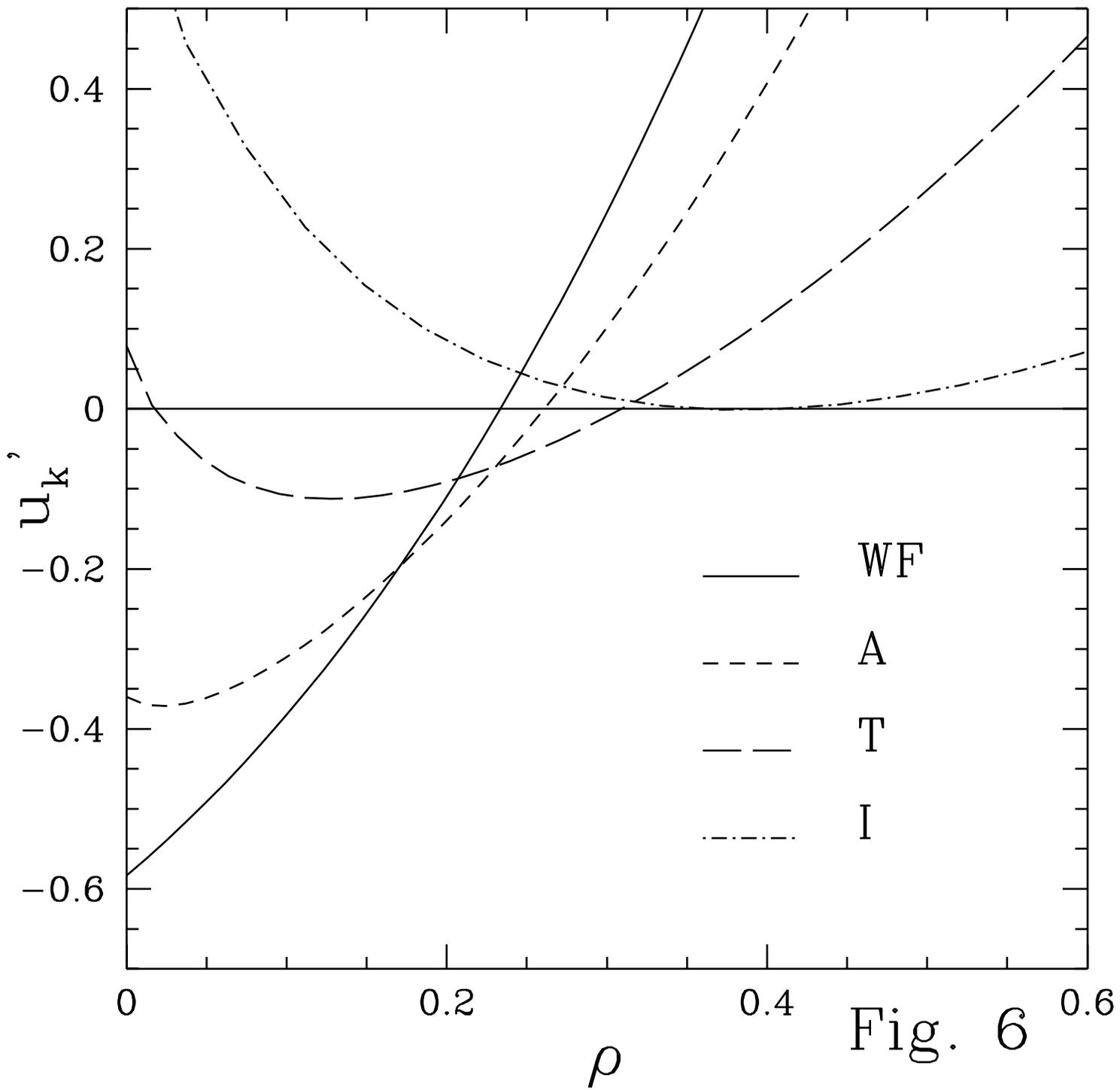
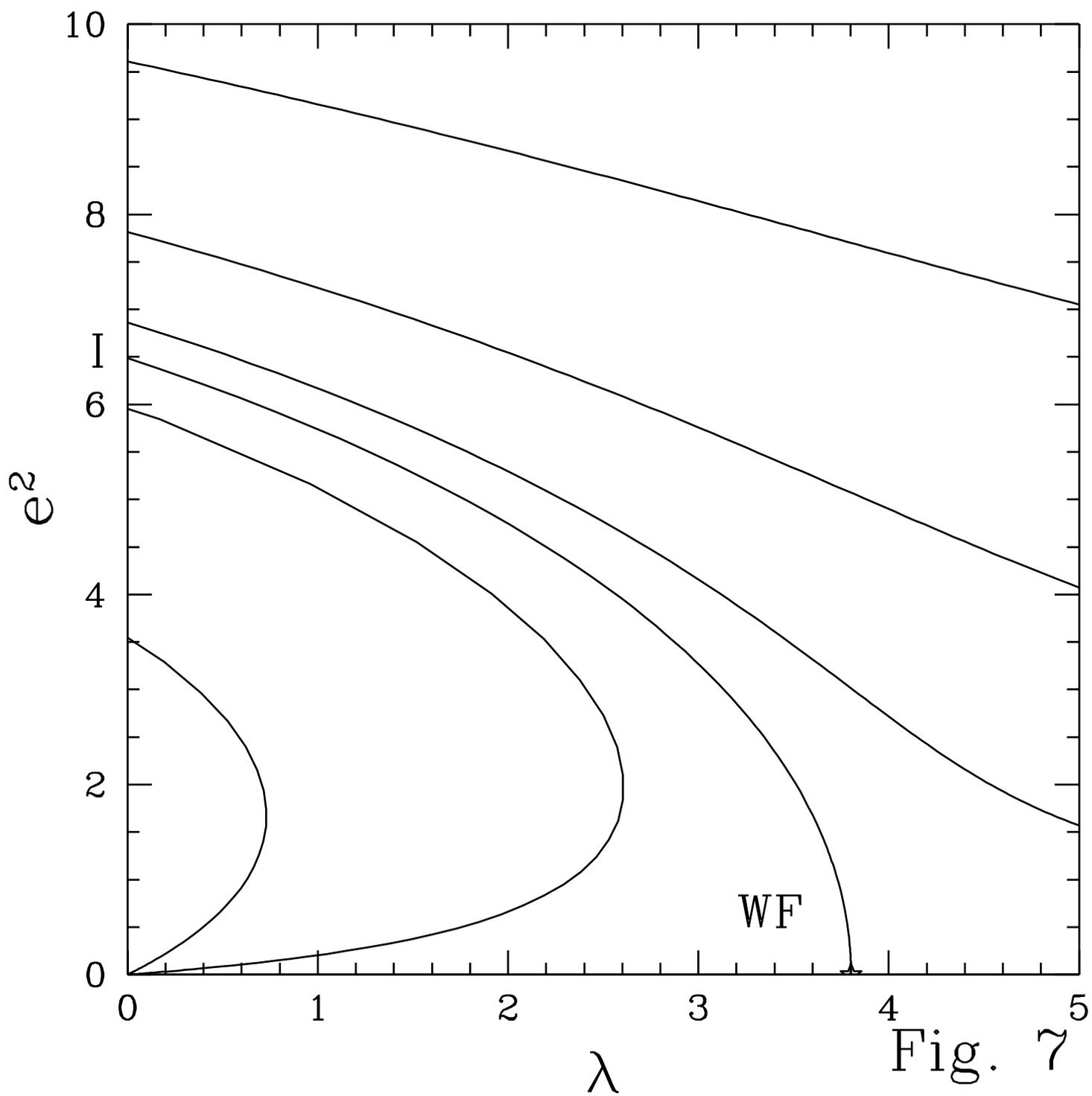
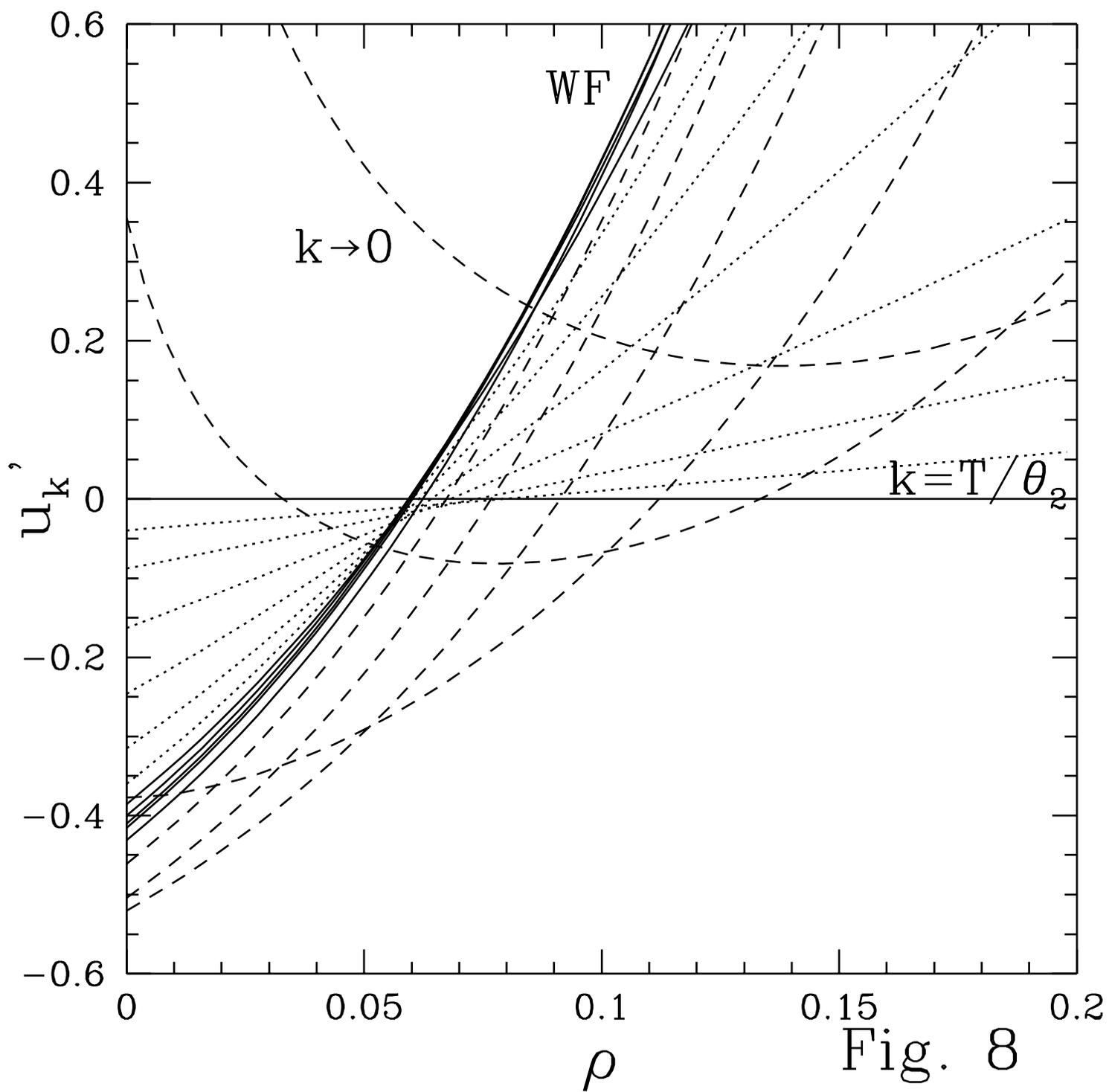
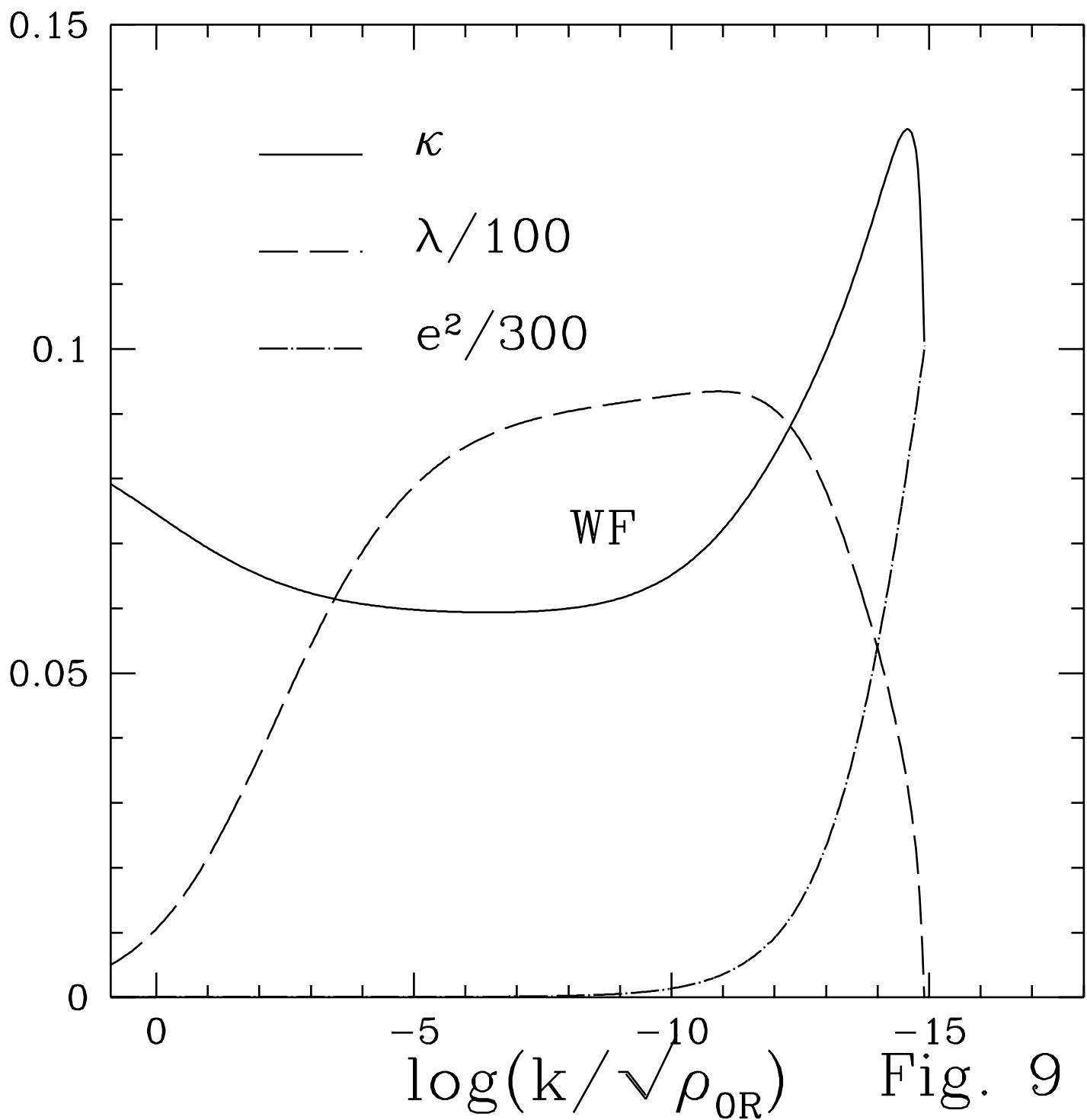
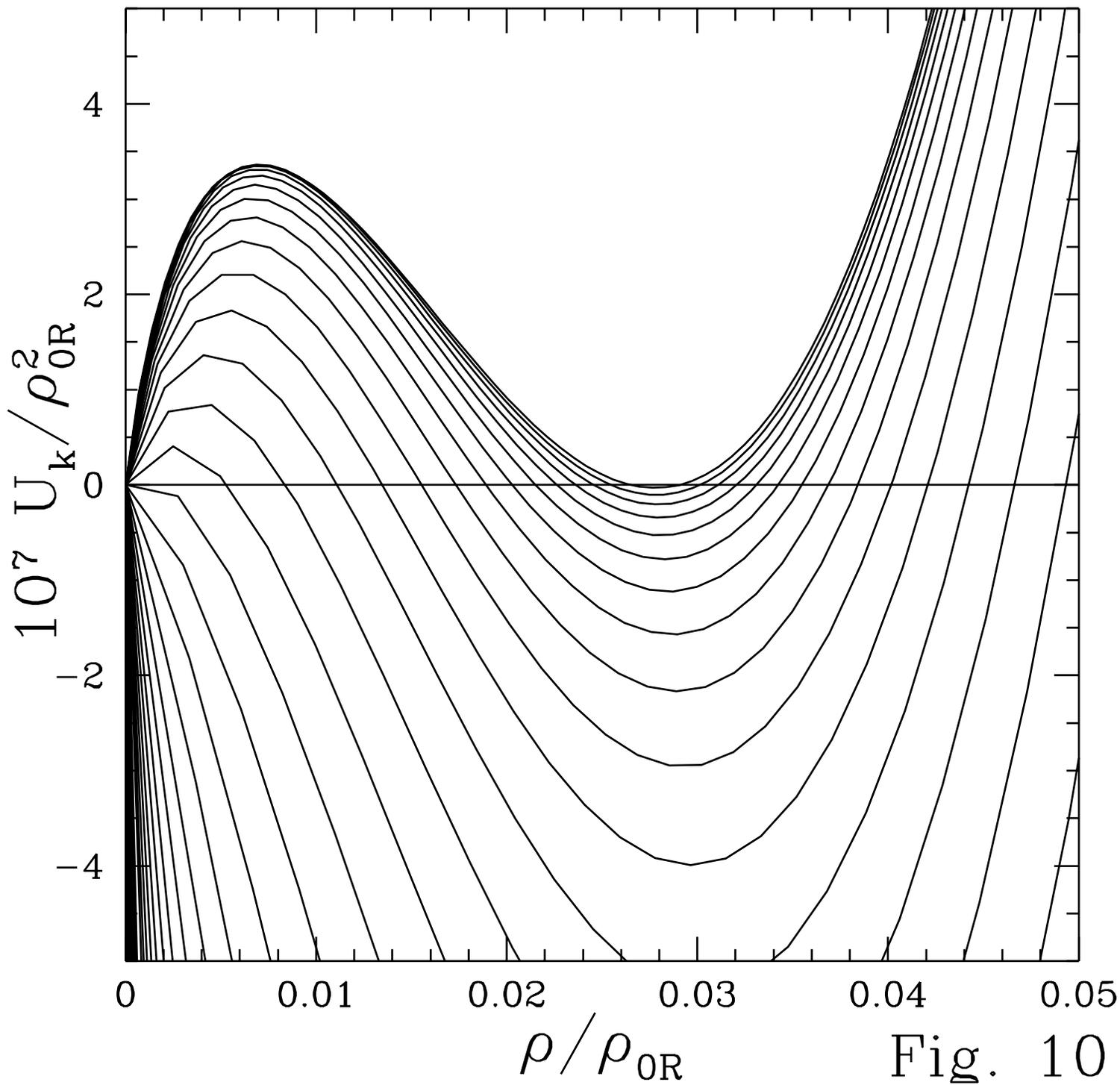
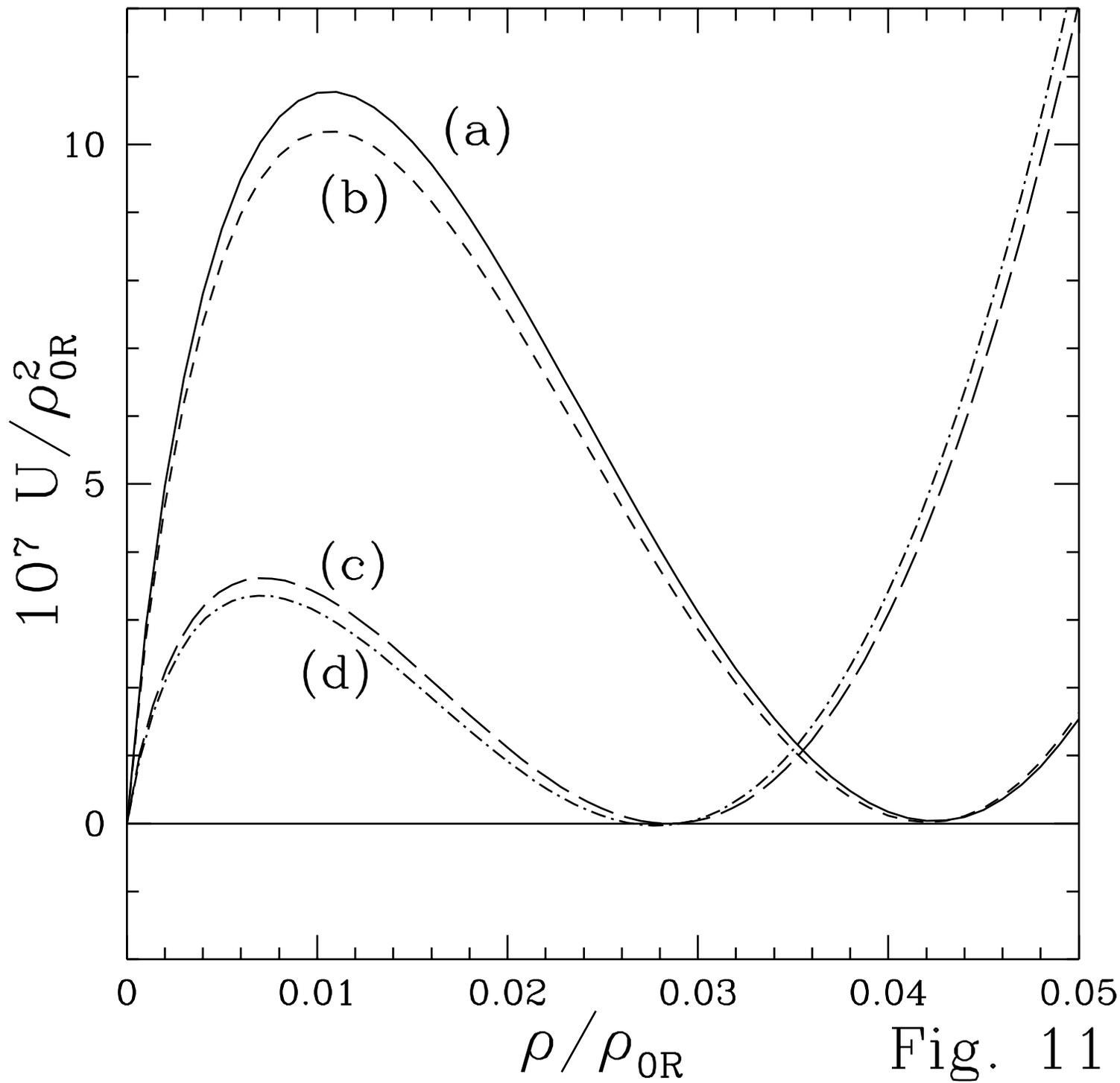
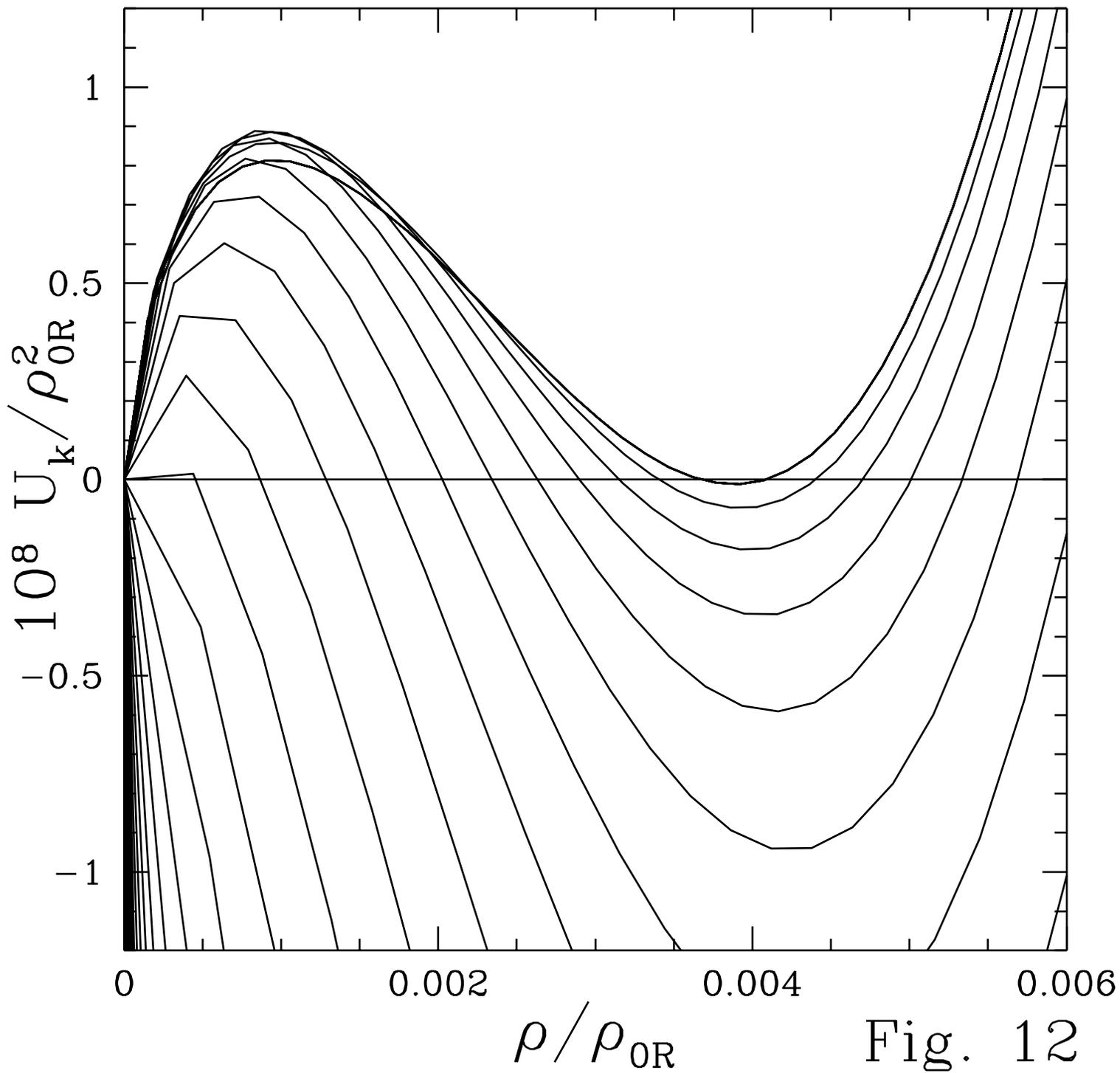
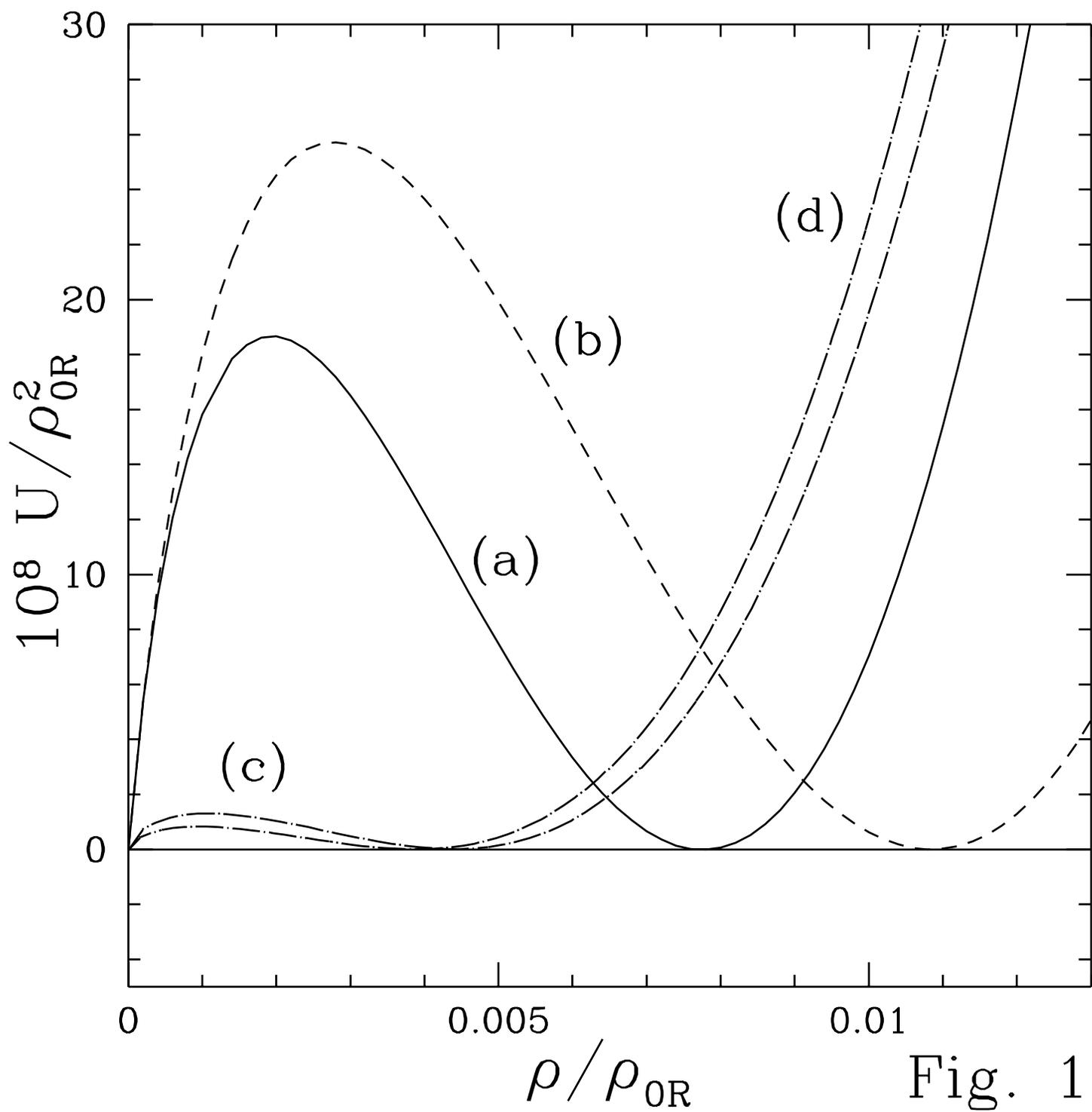
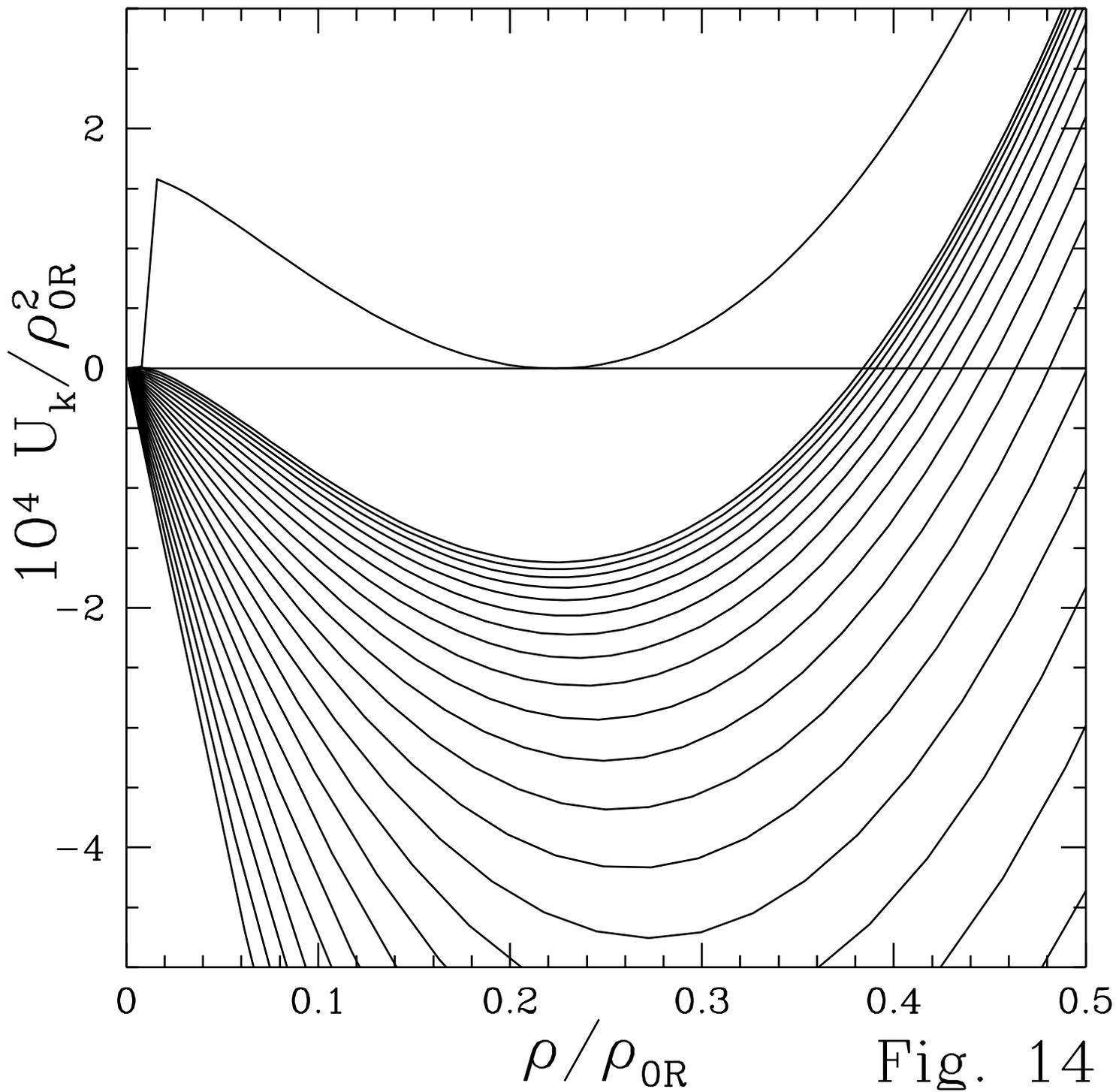
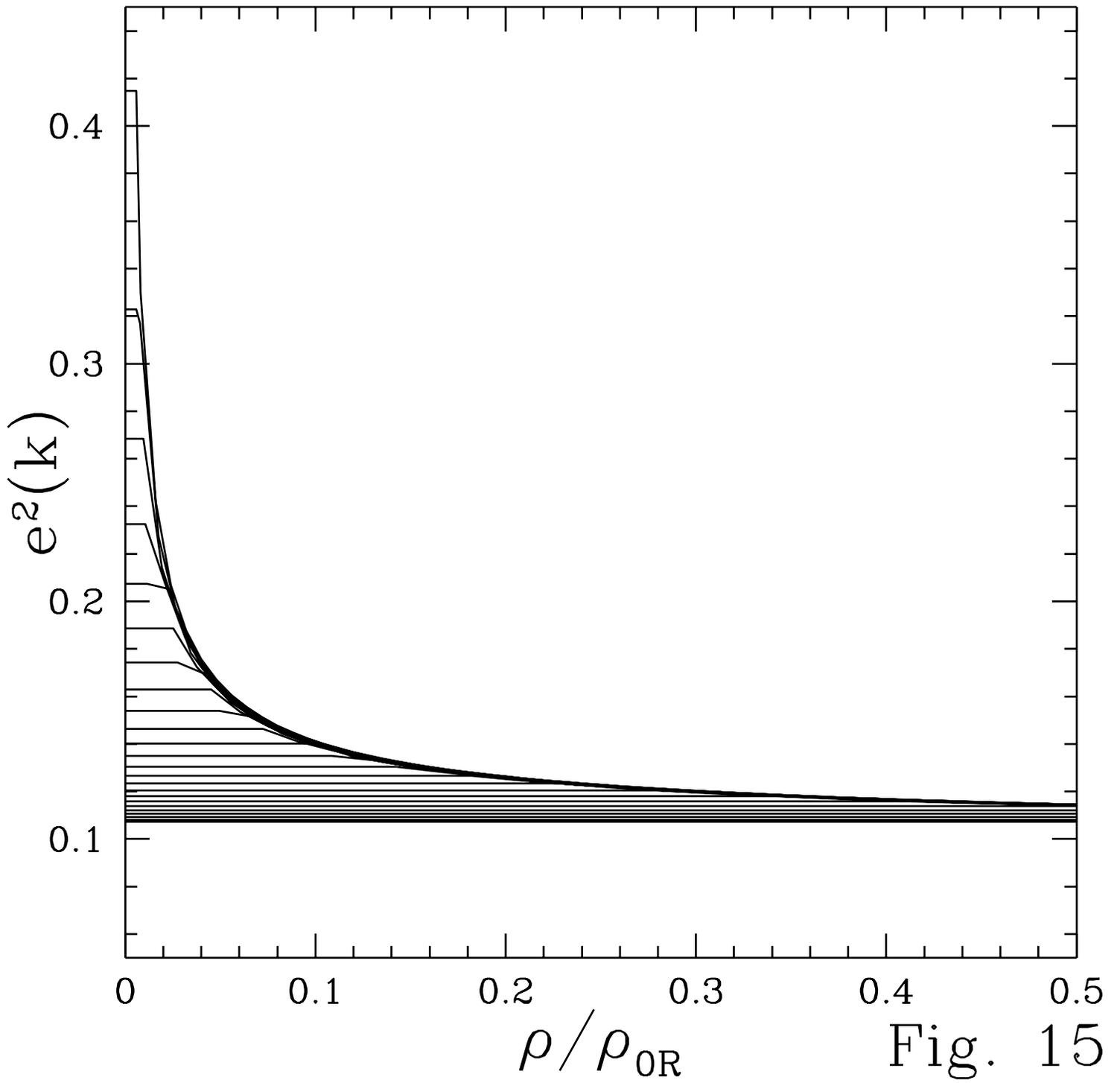
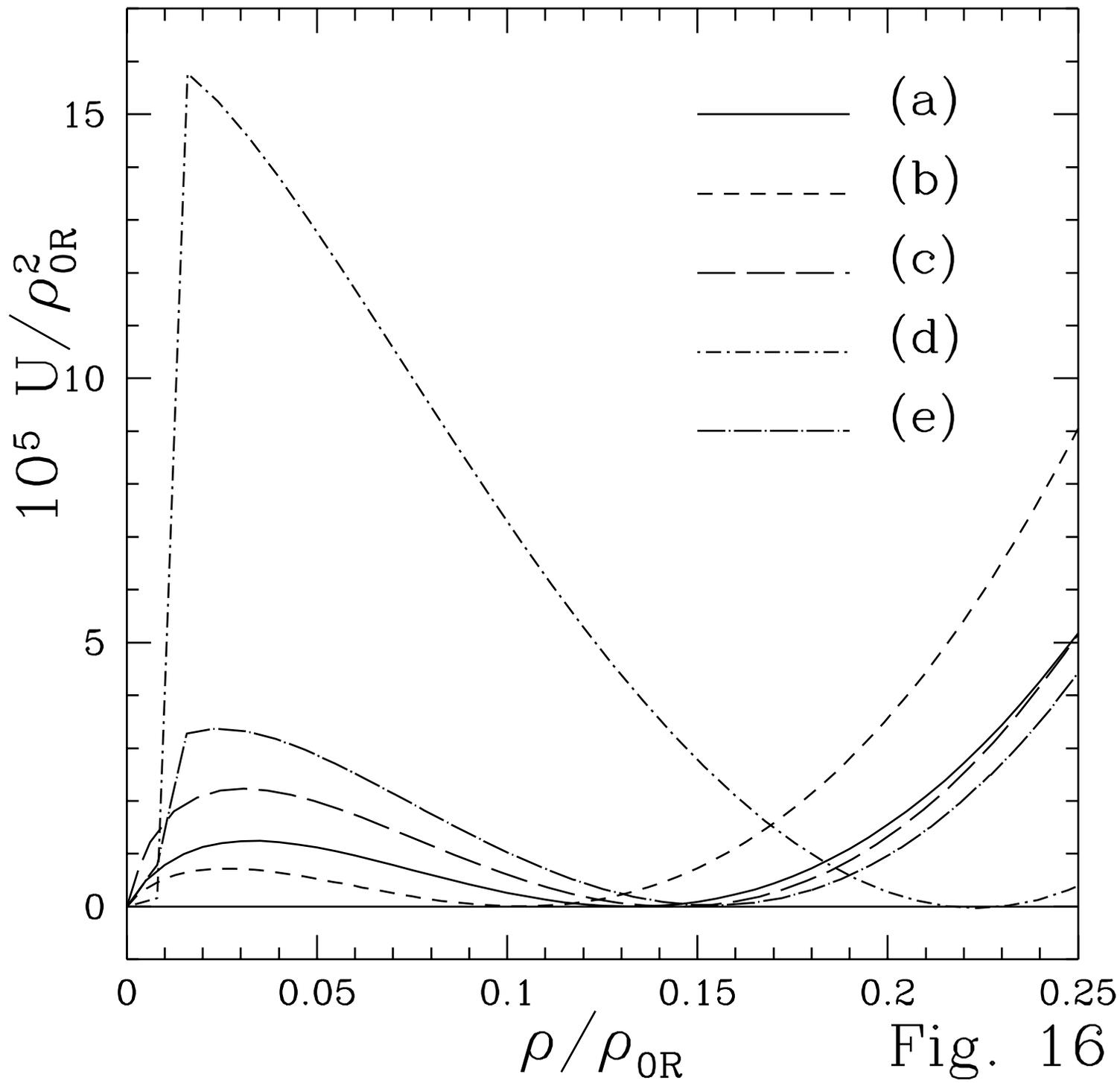
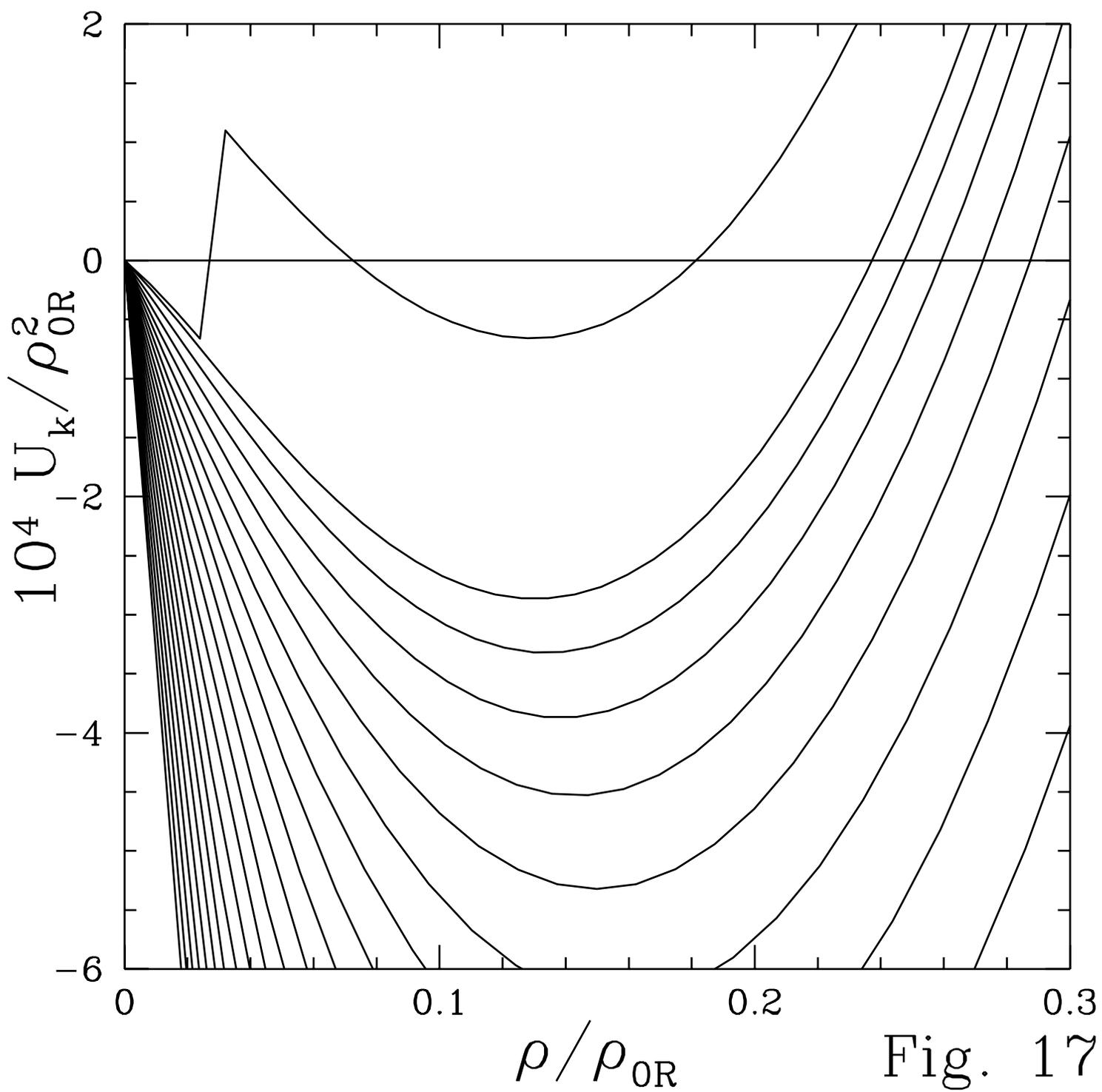
\end{document}